\documentclass[a4paper]{aa} % for the final version
\usepackage{epsfig,natbib,pjournal}

\begin{document}

%   \thesaurus{     12 (
%         11.05.2; % Galaxies: evolution
%         11.16.1; % Galaxies: photometry
%         12.03.3; % Cosmology: observations
%         13.09.1; % Infrared: galaxies
%         04.19.1) % Surveys
%         }

\def\F{FIRBACK }
\def\mic{\,$\mu m$}

\title{FIRBACK: III. Catalog, Source Counts, and Cosmological
Implications of the 170\mic \, 
ISO\thanks{Based on observations with ISO, an ESA project with instruments
funded by ESA Member States (especially the PI countries: France,
Germany, the Netherlands and the United Kingdom) and with the participation of ISAS and
NASA.}
 Deep Survey}
\subtitle{}
\titlerunning{FIRBACK III. Deep 170\mic\, ISO survey: Catalog, Source Counts \&
Implications.}
%\authorrunning{H. Dole {\it et al.}}
%________________________________________________________________
\author{Herv\'e Dole\inst{1,2} \and 
     Richard Gispert\inst{1} \and 
     Guilaine Lagache\inst{1} \and 
     Jean-Loup Puget\inst{1} \and
     Fran\c cois R. Bouchet\inst{3} \and
     Catherine Cesarsky\inst{4} \and
     Paolo Ciliegi\inst{5} \and
     David L. Clements\inst{6} \and
     Michel Dennefeld\inst{3} \and 
     Fran\c cois-Xavier D\'esert\inst{7} \and
     David Elbaz\inst{8,9} \and
     Alberto Franceschini\inst{10} \and
     Bruno Guiderdoni\inst{3} \and
     Martin Harwit\inst{11} \and
     Dietrich Lemke\inst{12} \and
     Alan F. M. Moorwood\inst{4} \and
     Sebastian Oliver\inst{13,14} \and
     Willian T. Reach\inst{15} \and
     Michael Rowan-Robinson\inst{13} \and
     Manfred Stickel\inst{12}
}
%________________________________________________________________
%\offprints{Herv\'e Dole (email: hdole@as.arizona.edu {\it or} Herve.Dole@ias.fr)}
\mail{hdole@as.arizona.edu}% {\it or} Herve.Dole@ias.u-psud.fr}
%________________________________________________________________
\institute{
     $^1$ Institut d'Astrophysique Spatiale, b\^atiment 121, Universit\'e Paris Sud, F-91405 Orsay Cedex, France \\
     $^2$ {\it present address:} Steward Observatory, University of Arizona, 933 N Cherry Ave, Tucson, AZ 85721, USA \\
     $^3$ Institut d'Astrophysique de Paris, 98bis Bd Arago, F-75014 Paris, France\\
     $^4$ ESO, Karl-Schwarzschild-Strasse 2, D-85748 Garching bei M\"unchen, Germany\\
     $^5$ Osservatorio Astronomico di Bologna, Via Ranzani 1, I-40127 Bologna, Italy\\
     $^{6}$ Department of Physics \& Astronomy, Cardiff University, PO Box 813, Cardiff CF22 3YB, UK\\
     $^{7}$ Laboratoire d'Astrophysique, Obs. de Grenoble, B.P. 53, 414 av. de la piscine, F-38031 Grenoble cedex 9, France\\
     $^{8}$ Service d'Astrophysique, DAPNIA, DSM, CEA Saclay, F-91191, Gif-sur-Yvette, France\\
     $^{9}$ Physics and Astronomy Departments, University of California, Santa Cruz, CA 95064, USA\\
     $^{10}$ Astronomy Departement, Padova University, Vicolo Osservatorio, 5, I-35122 Padova, Italy\\
     $^{11}$ 511 H. Street S. W., Washington DC 20024-2725, USA; also Cornell University\\
     $^{12}$ Max-Planck-Institut f\"ur Astronomie, K\"onigstuhl 17, D-69117 Heidelberg, Germany\\
     $^{13}$ Astrophysics Group, Imperial College, Blackett Laboratory, Prince Consort Road, London SW7 2BZ, UK\\
     $^{14}$ Astronomy Centre, University of Sussex, Falmer, Brighton BN1 9QJ, UK\\
     $^{15}$ IPAC, California Institute of Technology, Pasadena, CA 91125, USA
}
%________________________________________________________________
\date{
Received 22-Jan-2001; accepted 9-March-2001.} % v20010314am XXX
%________________________________________________________________
\abstract{
The FIRBACK (Far Infrared BACKground) survey is one of the deepest imaging surveys
carried out at 170 $\mu m$ with ISOPHOT onboard ISO, and is aimed at the study of
the structure of the Cosmic Far Infrared Background.
This paper provides the analysis of resolved sources. After a validated process of
data reduction and calibration, we perform intensive simulations to optimize
the source extraction, measure the confusion noise ($\sigma_c = 45$ mJy),
and give the photometric and astrometric accuracies. 196 galaxies with flux
$S > 3 \sigma_c$
are detected in the area of 3.89 square degrees.
Counts of sources with flux
$S > 4 \sigma_c$
present a steep slope of
$3.3 \pm 0.6$
on a differential "logN-logS" plot between 180 and 500 mJy.
As a consequence, the confusion level is high and will impact
dramatically on future IR deep surveys.
This strong evolution, compared with a slope of 2.5 from Euclidian geometry,
is in line with models implying a strongly evolving Luminous Infrared Galaxy population.
The resolved sources account for less than 10\%
of the Cosmic Infrared Background at 170 $\mu m$,
which is expected to be resolved into sources in the 1 to 10 mJy range.
}
\maketitle
%
%________________________________________________________________
\section{Introduction}
The European Space Agency's Infrared Space Telescope, ISO
\cite[]{kessler96,kessler2000} performed about 1000 programs between 1995
and 1998, including the deepest extragalactic observations ever made
in the mid- and far-infrared range with an unprecedented sensitivity
(for a review see \cite{genzel2000}).
Most of these deep cosmological observations aim at probing 
galaxy formation and evolution, mainly by resolving the Cosmic Infrared
Background (CIB) into discrete sources, but also by studying the CIB
fluctuations.

Understanding and observing the sources contributing to the
extragalactic background at all
wavelengths has become one of the most rapidly evolving fields in observational
cosmology since the discovery of the CIB \cite[]{desert95,puget96}.
In particular, deep observations from space with ISO, and from the ground
with SCUBA on the JCMT and MAMBO on the IRAM 30m telescope, respectively in the
infrared, submillimeter and millimeter
range, together with observations at other wavelengths for source
identification (in the radio and optical / NIR range), begin to provide a global view  
of galaxy evolution. The long wavelength observations reveal galaxies through their
dust emission, providing a complementary and significantly different view to that of optical 
and UV observations.

The ISO legacy regarding galaxy evolution includes a number of significant studies.
About a dozen deep surveys have been conducted in the mid infrared
with ISOCAM \cite[]{cesarsky96}, reaching
sensitivity levels of $30 \, \mu Jy$ at 15 \mic \,
\cite[]{altieri99,elbaz99,aussel99,desert99,flores99}.
The major results of the mid-infrared surveys involve source
counts obtained by combining a number of surveys. These exhibit
strong evolution with a steep slope up to $2.4 \pm
0.2$ \cite[]{elbaz99} in the integral logN-logS diagram. 
Multiwavelength identifications and
redshift distributions
constrain the nature of the sources \cite[]{flores99,aussel99,chary2001}:
most of them are Luminous Infrared Galaxies, LIRG's, 
at a median redshift of 0.8.

In the far-infrared, the 60 - 240 \mic \, spectral domain was
explored using the imaging capabilities of ISOPHOT (PHT) \cite[]{lemke96}.
As indicated in Figure 1 of \cite{gispert2000}, this domain
corresponds to the maximum emission of the extragalactic
background .
The main surveys published were carried out in the
Lockman Hole on $1.1$ sq. deg. at 90 and 170 \mic \, by
\cite{kawara98}, in the \F Marano field at 170 \mic \, by \cite{puget99}
and in the entire \F survey by \cite{dole99}, in SA57 on $0.4$ sq. deg. at 60 and 90 \mic \, by
\cite{linden-voernle2000}, and
in 8 small fields covering nearly $1.5$ sq. deg. at
90, 120, 150 and 180 \mic \,  by \cite{juvela2000}. A shallower
survey was performed over an area of $11.6$ sq. deg. at 90 \mic \, 
by \cite{efstathiou2000} as part of the ELAIS survey. The
ISOPHOT Serendipity Survey at 170 \mic \,
\cite[]{stickel98,stickel2000}
took advantage of ISO slews between targets to detect about
1000 sources between 1 and 1000 Jy.

\begin{figure}[!ht]
     \begin{center}
     \epsfxsize=9cm
     \epsfbox{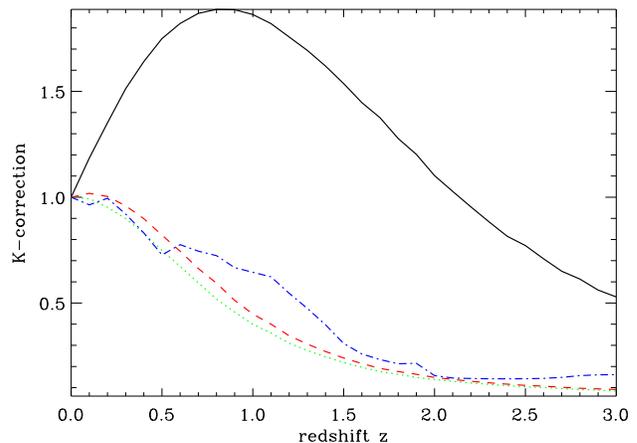}
     \caption[]{\label{fig:k-corrections} K-corrections 
at 15 (dot-dashed curve), 60 (dotted curve), 90 (dashed curve) and 170 \mic \,(solid curve) for a
LIRG
\cite[]{dole2000b,lagache2001b}. The wavelengths of cosmological interest
are thus around 15 \mic\, and above 150\mic\, where they benefit from
the ``negative K-correction effect'', increasing the sensitivity up to
redshifts around unity.}
     \end{center}
\end{figure}

In the 60 to 120 \mic \, spectral windows, the
\verb+C_100+ camera, with its $3 \times 3$ array of Ge:Ga detectors, was subject to strong
transients and spontaneous spiking, limiting the sensitivity (which is
a few times better than IRAS); fortunately, new attemps to overcome
these problems with a physical model of the detector seem promising
(Coulais et al., 2000; Lari \& Rodighiero, 2001).  At 60 and 90 \mic, no clear evolution in the
source counts is observed, since both non-evolution and
moderate evolution models can still fit the data \cite[]{linden-voernle2000,efstathiou2000}. 
Furthermore, the K-correction\footnote{K-correction is defined as
the ratio: $\frac{L(\nu ')}{L(\nu)}$ where $L(\nu)$ is the luminosity
at frequency $\nu$, and $\nu = (1+z) \nu '$. Thus, $K(z) =
\frac{L(\nu \times [1 + z])}{L(\nu)}$.}
(Figure \ref{fig:k-corrections}
from the model of \cite{dole2000b} and \cite{lagache2001b}) between 30 and 120 \mic \, is not
favorable for probing galaxy evolution up to redshifts $z\sim 1$.
With the ELAIS survey, \cite{serjeant2000a} were able to
derive the luminosity function of galaxies up to redshift $z \simeq 0.3$.

%________________________________________________________________
% FIGURE K-corrections

At longer wavelengths (120-240 \mic), the \verb+C_200+ camera,
a $2 \times 2$ array of stressed Ge:Ga detectors, is more stable and
most of the detectors'  behaviour can be characterized and, if needed,
properly corrected \cite[]{lagache2001}.  
The K-correction at 170 \mic \, (Figure \ref{fig:k-corrections}),
as well as in the mid-infrared around 15 \mic,
is favorable and becomes optimal at redshifts around 0.7.
The first analysis of deep surveys at 170 \mic \, showed a large excess in source counts
over predictions of no-evolution models 
at flux levels below 200 mJy \cite[]{kawara98,puget99}, suggesting strong
evolution. Recent work by \cite{juvela2000} is in agreement with this
picture, and includes the far-infrared colors of the sources.

The \F survey (acronym for Far Infrared BACKground) 
was designed to broaden our understanding of galaxy evolution
with its accurate source counts and its catalog allowing
multiwavelength follow-up.  It also enabled studies of
the CIB fluctuations (first detected in the first area surveyed
in the FIRBACK program by \cite{lagache2000a}).
\F is one of the deepest surveys made at 170 \mic\, and the
largest at this depth. This survey used about 150h of observing time, corresponding to the
$8^{th}$ largest ISO
program \cite[]{kessler2000}. 

The aim of this paper is to provide the catalogs and the source counts
of the \F survey. Preliminary \F source counts were published by
\cite{lagache98} and \cite{puget99} on the $0.25$ sq. deg. Marano 1
field, and by \cite{dole2000} on the entire survey.
An overview of this paper is as follows. Section 2 presents the
observational issues of the \F survey and Section 3 summarizes
the data processing and the calibration (a complete description can be found
in \cite{lagache2001}). Section 4 explains the
extensive simulations and the source extraction technique.
Section 5 details the flux measurement by aperture photometry,
analyses the photometric and astrometric noise of the sources and provides estimates of
accuracies.
In section 6 we present the final \F catalog ($S > 4\sigma_s$),
and the complementary catalog ($3\sigma_s < S < 4\sigma_s $)
extracted for follow-up purposes. Section 7 describes the corrections that have been
applied (completeness, Malmquist-Eddington effect) and presents the
final \F source counts at 170 \mic .
Section 8 compares our results to other observations as well as models,
and discusses the cosmological implications of the \F source counts:
strong evolution and resolution of the CIB.
%
%________________________________________________________________
\section{The \F survey: Fields \& Observations}
%
%________________________________________________________________
\subsection{Fields}
\F is a survey at 170\mic \, covering four square degrees in three
high galactic latitude fields, called \F South Marano (FSM), \F North
1 (FN1) and \F / ELAIS North 2 (FN2) (see Table~\ref{tab:fields}). They were
chosen to have foreground contaminations as low as possible: the
typical HI column-density is less than or equal to $N_H \simeq 10^{20}
cm^{-2}$, and the 100\mic\, brightness is less than 1.7 MJy/sr on DIRBE
maps. In addition, FN1 and FN2 were chosen to match some fields from
the European Large Area ISO Survey, ELAIS \cite[]{oliver2000}, which
had been covered at 15\mic\, with ISOCAM \cite[]{serjeant2000} and at
90\mic\, with ISOPHOT \cite[]{efstathiou2000}. FN2 observation time is
a collaboration between the ELAIS and \F consortia.
%
%________________________________________________________________
% TABLE FIELDS
\begin{table}[htb]
  \caption{\em Fields of the \F survey at 170\mic}
 \label{tab:fields}
  \begin{center}
    \leavevmode
    \footnotesize
    \begin{tabular}[h]{lccccc}
      \hline \\[-5pt]
      field    & $\alpha_{2000}$    &  $\delta_{2000}$   &  {\it l} & {\it b} & $S_{100}$$^a$\\[+5pt]
      \hline \\[-5pt]
      FSM      & 03$^h$ 11$^m$ & -54$^{\circ}$ 45' & 270$^{\circ}$ & -52$^{\circ}$ & 1.42 \\
      FN1      & 16$^h$ 11$^m$ & +54$^{\circ}$ 25' &  84$^{\circ}$ & +45$^{\circ}$ & 1.17 \\
      FN2      & 16$^h$ 36$^m$ & +41$^{\circ}$ 05' &  65$^{\circ}$ & +42$^{\circ}$ & 1.19 \\
      \hline \\
      \end{tabular}
    \begin{tabular}[h]{l}
	$^a$ Mean brightness at $100 \mu m$ (MJy/sr) in DIRBE maps\\
	(annual average, zodiacal component subtracted)\\
    \end{tabular}
  \end{center}
\end{table}
%
%________________________________________________________________
% TABLE OBSERVATIONS
\begin{table}[htb]
  \caption{\em Observational characteristics of the \F fields}
 \label{tab:obs}
  \begin{center}
    \leavevmode
    \footnotesize
    \begin{tabular}[h]{lccc}
      \hline \\[-5pt]
          field 		& FSM1    &  FN1    &  FN2  \\[+5pt]
      \hline \\[-5pt]
	area (sq. deg.)		& 0.95    & 1.98    & 0.96 \\ 
	rasters$^b$		& 4  & 2  & 2 \\
	redundancy$^c$		& 16 & 8  & 8 \\    
	$t_{int}$$^d$ (sec)	& 256     & 128     & 128 \\
	raster step$^e$ (pixels)& 1,1,    & 1,1     & 1,1 \\
				& 2,2$^a$ &    &    \\
	offset$^f$ (pixels)	& 0.5,0.5      & $<1$$^g$ & $<1$$^g$ \\
				& 1,1$^a$      &    & \\
	date			& Nov-1997     & Dec-1997     & Jan-1998\\
				& Jul-1997$^a$      &         &    \\
	revolution$^h$		& 739 to 744   &753 to 774    & 785 to 798\\
				& 593$^a$ &         &    \\
      \hline \\
      \end{tabular}
    \begin{tabular}[h]{l}
     $^a$ in the case of the FSM1 field only\\
     $^b$ number of different rasters mapping the same field\\
     $^c$ number of different observations per sky pixel on the \\center
      of final coadded map\\
     $^d$ integration time per sky pixel on the center \\
      of final coadded map \\
     $^e$ offset in pixel in the Y and Z directions of the spacecraft\\
     between the steps on the raster\\
     $^f$ offset in pixel between different rasters\\
     $^g$ offset is irregular due to the rotation of the fields\\
     $^h$ ISO revolution numbers (or number range) of observation\\
    \end{tabular}
  \end{center}
\end{table}
\vspace{-1cm} % XXX
%
%________________________________________________________________
\subsection{Observations}
Observations were carried with ISO, using the ISOPHOT spectro-photo-polarimeter. We used
the \verb+C_200+ $2 \times 2$ pixel photometer and \verb+C_160+ broadband filter centered at
$\lambda = 170\, \mu m$. Scanning the sky was done in raster map mode, AOT P22, with one
pixel offset between each pointing, to provide the redundancy. Individual rasters were shifted
with respect to each other by a fraction of a pixel to provide proper
sampling where possible.
Table \ref{tab:obs} summarizes the observational characteristics of the fields.

The FSM field is composed, for historical reasons, of four individual
fields, called FSM1, 2, 3 and 4 (Figure 6 in \cite{lagache2001}).
FSM1 on the one hand, and FSM2, 3 and 4 on the other, have been observed continuously:
transient effects are thus reduced and no rotation of the field occurs between different rasters
(same roll angle).
FSM1 rasters are offset by two pixels in order to maximise redundancy and establish the
ISOPHOT sensitivity for such observations, whereas FSM2, 3 and 4 are offset by a half pixel in
both Y and Z directions to increase oversampling. 

The FN1 field is composed of eleven individual fields (Figure 7 in \cite{lagache2001}),
observed twice. 
Observations were not performed continuously, so that each individual raster has a different roll
angle, giving a sampling of the sky that is non uniform.

The FN2 field is composed of nine individual fields (Figure 8 in \cite{lagache2001}), 
observed twice. The other characteristics are the same as for FN1.
%
%________________________________________________________________
\section{Data Reduction, Instrumental Effects, Calibration, Maps}
\label{datareduction}
The complete process of data reduction and calibration is described in
\cite{lagache2001}. Here, we merely summarize the different steps.
%
%________________________________________________________________
\subsection{Interactive Analysis}
We made use of the PHT Interactive Analysis package (PIA) version 7.2.2 \cite[]{gabriel97} in
the IDL version 5.1 environment, to process the raw data (named ERD: Edited Raw Data) into
brightnesses (named AAP: Astronomical and Application product). After linearizing and
deglitching the ramps, we applied the orbit-dependent dark and reset interval corrections. We
calibrated the data with the two bracketing FCS lamps (Fine Calibration Source) values, using the mean
value in  order not to induce baseline effects.
%
%________________________________________________________________
\subsection{Glitches, Long Term Transients, Flat Fielding}
Cosmic particles hitting the detector are easy to detect at the time
of their impact, but they may cause response variations.
On 224 different measurements (that is 56 independent rasters observed by 4 pixels), we report
only 13 such cases, which are corrected. Furthermore, thanks to the high redundancy of each
raster, a glitch cannot mimic a source because the same piece of the sky is observed
independently by the four pixels of the photometer at different times.

Some long term transients (LTT) are seen in the data, and are understood to be the consequence
of step fluxes seen by the photometer. During the FIRBACK observations, ISOPHOT was looking at
relatively flat fields with low background, but was on more complex fields
during the preceding observations. Our
best data occur where the observations were made continuously. 
We correct for the LTT by forcing all the pixels to follow the time variations of the most stable
pixel, which is assumed to represent the sky. This correction is found to be linear, and never
exceeds 10 \%.
%The sophisticated LTT correction method developed by \cite{miville-deschenes2000} for
%ISOCAM data with high redundancy (thanks to the $32 \times 32$ array), is not efficient in our
%case because of insufficient redundancy with the $2 \times 2$ pixel camera.

We then compute a flat field using the redundancy and apply the necessary corrections. The
detector
behaviour is highly reproducible, leading to constant 
flat field values: $1.04 \pm 0.02$, $0.91 \pm 0.02$,
$1.09 \pm 0.02$
and $ 0.94 \pm 0.02$
for pixels 1, 2, 3 and 4 respectively.

%
%________________________________________________________________
\subsection{Photometric Correction}
There is a difference of 11\% between the solid angle value of the PHT footprint 
at 170\mic\, used by PIA and the value derived by calibration
observations around Saturn and the model.
We thus apply a multiplicative correcting factor of 0.89 to the brightness values given by PIA to take into
account the real profile of the footprint.
%
%________________________________________________________________
\subsection{Maps}
For a given raster measurement, we project the signal from each pixel on a regular grid
defined by the raster. Between each pointing, we make an interpolation and check that the
photometry is not changed by more than 1\%.
Then we sum all these signals 
on a celestial coordinate grid to get the final map.
%
%________________________________________________________________
\subsection{Calibration of Extended Emission}
Using the knowledge of the average interstallar dust emission
spectrum, the zodiacal light emission at the
time of the observations, and the Cosmic Infrared Background values derived from COBE,
together with HI data on our fields, we derive a brightness value
at 170\mic\, for each of our fields.
This extrapolated brightness at 170 $\mu$m for the three fields is in remarkable
agreement with the measured ISOPHOT brightness.
Furthermore, the rejection level of straylight up to $60^{\circ}$ off-axis observed by ISO during
total solar eclipse by the Earth, is better than $10^{-13}$, implying that there is no significant
contribution to the measured flux coming from the far sidelobes. This confirms that ISO is able
to make absolute measurements of the extended emission and gives a high degree of confidence
to our photometric calibration.
%
%________________________________________________________________
\section{Source Extraction, Simulations}
\label{simulations}
An important part of the present 
work is the extraction of the sources, the simulation of point source observations and the analyses
of noise. After detecting sources on a median-filtered-like map, we measure the fluxes on the
final maps with aperture photometry. Our simulation tool validates the flux determination as well
as the noise analysis.
%
%________________________________________________________________
\subsection{Source Extraction}
Our original maps are dominated by the fluctuations of the background
at 170\mic, at all spatial scales, mainly due to the cirrus
confusion noise and the CIB fluctuations \cite[]{lagache2000a}. Because of this, classical
extraction algorithms based on thresholding and local
background determination mostly fail: it is not easy to use a robust
detection algorithm on maps dominated by structures at all scales. On
the contrary, flat background maps allow reliable detection with
the available processing techniques, like gaussian fitting methods,
e.g. for faint ISOCAM sources by \cite{desert99}. Because of the
undersampling of the PHT Point Spread Function together with a highly
fluctuating background, CLEAN-like methods \cite[]{hogbom74} are
difficult to use. Wavelet decomposition, e.g. for ISOCAM by
\cite{starck99}, is not easily implementable because of the poor spatial
dynamics of our maps (``big pixels and small maps'').
To overcome these difficulties we have developed the following method by
combining some well-known techniques for source extraction and flux
determination:
%________________________________________________________________
% FIGURE SOURCE MAP
\begin{figure}
     \begin{center}
     \epsfxsize=8cm
     \epsfbox{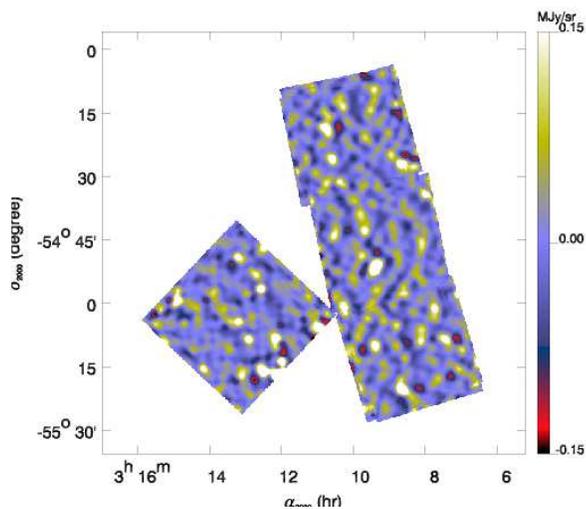} 
     \caption[]{\label{fig:source_map}Example
of a {\it source map} for source detection in the FSM
field. Background is subtracted using a 
median filter in the time space (AAP). Data with only high spatial
frequencies are then reprojected on a map
with the FIRBACK pipeline.}
     \end{center}
\end{figure}
\begin{itemize}
\item[$\bullet$] background is subtracted in the one dimensional time
data (AAP level, brightness as a function of time) using a
median filter (size: 5 positions) to create {\it source time data}
\item[$\bullet$] {\it source time data} are processed to create 2-dimensional {\it source maps}
(Figure~\ref{fig:source_map}) through the FIRBACK pipeline as decribed
in Section \ref{datareduction}
\item[$\bullet$] source detection is performed on the {\it source
maps} using SExtractor \cite[]{bertin96}
\item[$\bullet$] flux measurements are performed on the unfiltered maps, 
using aperture photometry at the positions found by the source
detection only if there are at least 4 different observations,
and make a temporary version of the source catalog
\item[$\bullet$] by subtracting iteratively the brightest sources from the
temporary catalog using a CLEAN-like method on the final maps, we
remeasure with better accuracy the flux of the sources which have bright neighbours. This gives
the final catalog after two more corrections: short term transient of 10\%,
and flux offset of about 15\% derived from simultation (see Section \ref{photometry}).
\end{itemize}
Source detection is performed using SExtractor version 2.1.0 on the {\it source maps} with the
parameters given in Table~\ref{tab:sex_params}. Note that we do not use the background
estimator and set it to a constant value because {\it source maps} are flat maps containing
fluctuations due to resolved sources, since the background has been filtered.
Only the positions in the map of the detected sources will be used in
the output catalog computed by SExtractor (e.g. not the flux). We discard the edges by considering only
parts of the sky that have been observed at least 4 times. This reduces the total area by about 5\%.
%________________________________________________________________
% TABLE SEXTRACTOR PARAMS
\begin{table}
  \caption{\em Parameters used in SExtractor 2.1.0 applied on the Source Maps}
 \label{tab:sex_params}
  \begin{center}
    \leavevmode
    \footnotesize
    \begin{tabular}[h]{cc}
      \hline \\[-5pt]
      Parameter     &  Value  \\[+5pt]
      \hline \\[-5pt]
     \verb+DETECT_MINAREA+    &    10 \\
     \verb+DETECT_THRESH+     &    3.0 \\
     \verb+BACK_SIZE+    &    10 \\
     \verb+BACK_FILTERSIZE+   &    1,1 \\
     \verb+BACK_TYPE+    &    \verb+MANUAL+ \\
     \verb+BACK_VALUE+   &    -0.04,0.0 \\
      \hline \\
      \end{tabular}
  \end{center}
\end{table}
%
%________________________________________________________________
\subsection{Simulations}
We have developed a simulation tool of point sources in order to validate the flux determinations
and study source completeness of our survey. \cite{kawara98} did not make such simulations and
\cite{juvela2000} only tested the significance of their source detection because of a lack of redundancy 
in their observations. 
The work of \cite{efstathiou2000} included large simulations at 90 \mic, but the source detection is performed by eye. 

Thanks to the quiet behaviour of the \verb|C200| camera at 170 \mic, together with redundancy,
the detector noise as well as effects induced by
glitches can be neglected to first order with respect to the confusion
noise. (This is unlike conditions applying to the \verb|C100| camera 
\cite[]{linden-voernle2000}.)  

Here, we present a summary of our simulation process, followed by some details concerning the
addition of the sources and the validation:
\begin{itemize}
\item[$\bullet$] select a random sky position for a simulated source inside a \F field
\item[$\bullet$] add the source in each raster in AAP level which has observed the source itself
or its wings
\item[$\bullet$] process maps through the \F pipeline
\item[$\bullet$] extract sources with SExtractor
\item[$\bullet$] identify the extracted sources by comparing the
coordinates with the input catalog
\item[$\bullet$] compute a flux with aperture photometry using the effective footprint
(defined in Section \ref{fluxmeas})
\item[$\bullet$] validate on different flat backgrounds
\item[$\bullet$] validate on real data: different input fluxes and positions
\end{itemize}
%________________________________________________________________
% FIGURE SIMULATED SOURCES
\begin{figure}[!h]
     \begin{center}
     \epsfxsize=8cm
     \epsfbox{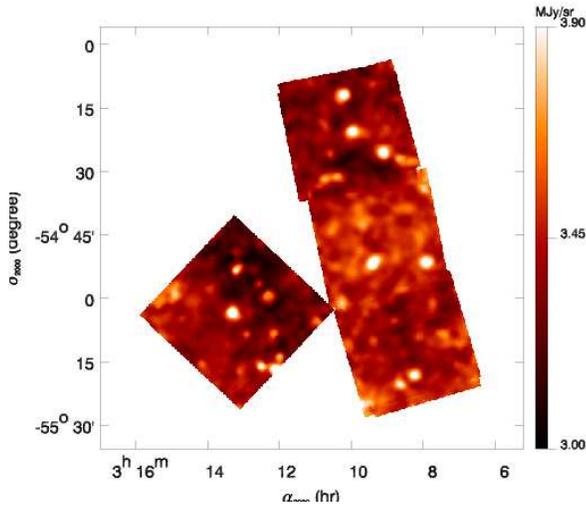}
     \caption[]{\label{fig:map_simu_Marano}Example of the addition
of 500 mJy sources in the FSM field. There are 8 sources spread
randomly throughout the field. One example is near the center of the
eastern survey square (FSM1).}
     \end{center}
\end{figure}
%
%________________________________________________________________
\subsubsection{Adding the Sources}
We use the best footprint available for PHT at 170\mic\, \cite[]{lagache2001} to simulate a
source with a known input flux; its spatial extension is taken to be a five pixel square, that is
about $7.7' \times 7.7'$ (note that the PIA footprint profile given in the calibration files extends
to only 4.2 arcminutes). This simulated source is added in the one dimensional time data (AAP
level). To avoid biases due to specific positions in the fields, we select random positions.

Because we have either 2 or 4 different raster observations of the
same parts of the sky, the randomly-selected sky position may fall e.g. on the edge
of a pixel in one raster, and at the center of another pixel in another
raster. We thus make the following approximation: we cut each PHT
pixel in 9 square sub-pixels of about $30.7 \times 30.7$ square
arcseconds. We compute the pixelized footprint for the nine
configurations corresponding to the cases where the source center
falls on one of the sub-pixels.

We make separate realizations for 8 input fluxes (100,
150, 200, 300, 500, 650, 800 and 1000 mJy) and create maps using the
\F pipeline. We add only between 6 and 20 sources per square degree at a time
depending on their flux, in order to avoid changing the confusion level when sources are added in the
data. We compute the needed number of maps to get 1200 realizations
for each flux in each field, or 28800 sources in total, in order to
have a statistically significant sample. We finally get about $2
\times 1230$ different simulated maps per field (1 final map + 1
source map for each realization) taking about 14 Gbyte, after about
one week of computation under IDL on a MIPS R12000 at 300MHz SGI.
Fig.~\ref{fig:map_simu_Marano} shows an example of added sources.

%
%________________________________________________________________
\subsubsection{Validation}
We extract sources on the final maps and compute fluxes as explained below by aperture
photometry. The aperture photometry filter parameters have been
optimized to obtain the best signal to noise ratio using the simulations. 

The validation is performed on flat background maps with different
surface brightness values (0.01, 3 and 10 MJy/sr), to check that the recovered flux
does not depend on the background. The difference between the input
and recovered flux is less than 1\% on an individual raster when the
source is centered on a pixel. When using random positions of the
sources and 2 or 4 rasters co-added, the recovered fluxes
have a dispersion explained by the ``edge effect'' (due to the
dilution of the flux in other pixels when the source falls on the
edge of a pixel)
and by the poor sampling of the sky, leading to an
overall uncertainty of 10\%. 
%
%________________________________________________________________
\section{Photometry, Noise Analysis, Accuracy}
\label{photometry}
%
%________________________________________________________________
\subsection{Flux Measurements by Aperture Photometry}
\label{fluxmeas}
Once a detection is obtained on source maps, fluxes have to be measured in final maps. 
Simulations of point sources on a flat background permit derivation of the
{\it effective average footprint} on the map, which results from the PHT
footprint and the final pixeling obtained in a given field, which depends
on the exact timing of the observations (roll angle).

\begin{figure}
     \begin{center}
     \epsfxsize=7cm
     \epsfbox{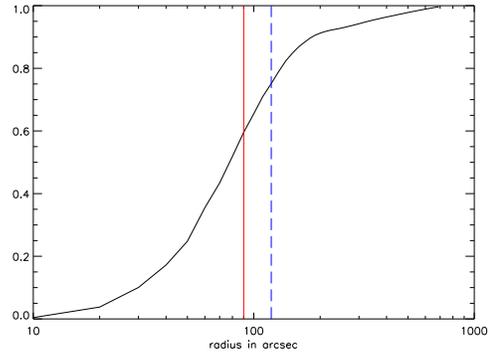}
     \caption[]{\label{fig:integral_psf_xl_ap}Growth curve of the
effective footprint on a logarithmic scale with the location of the radii of circles used for aperture
photometry; dotted vertical line: 90 arcsecond for the inner radius; dashed vertical line: 120
arcsecond for the outer radius.}
     \end{center}
\end{figure}

We check that strong sources in the data
have a profile in agreement with the effective footprint. The growth
curve of the effective footprint is plotted in Fig.~\ref{fig:integral_psf_xl_ap}.
The determination of the parameters for the aperture photometry filter
is performed by measurements of the flux of simulated sources through
different sets of apertures.
 
We find that the following values minimize the noise: an
internal radius of 90 arcseconds for measuring the source and an
external radius of 120 arcseconds to estimate the background. 
The determination of the flux takes into account the fact that at
these radii we select only a part of the effective footprint, and    
includes the appropriate correction. 

%________________________________________________________________
% FIGURE GROWTH CURVE AND FILTERS

In order not to be biased by a nearby strong source which could affect the estimate of the local
background in a measurement, we used a CLEAN-like procedure. We first compute a temporary
catalog that we sort by decreasing flux. Then we measure the brightest source, and remove it, and
repeat this process through the whole catalog. Note that this procedure is not used to extract faint
sources but only to improve the photometry of sources detected before applying the CLEAN procedure.

At the end of the process, we add 10\%
to the source flux to account for the transient behaviour of the
detector. This value is derived from our absolute measurement in the FSM1
(using AOT P25) in which the instantaneous response and the following
transient, as well as the final flux after 256 seconds, are observed
\cite[]{lagache2001}.
%
%________________________________________________________________
\subsection{Confusion Noise}
\begin{figure}
     \begin{center}
     \epsfxsize=7cm
     \epsfbox{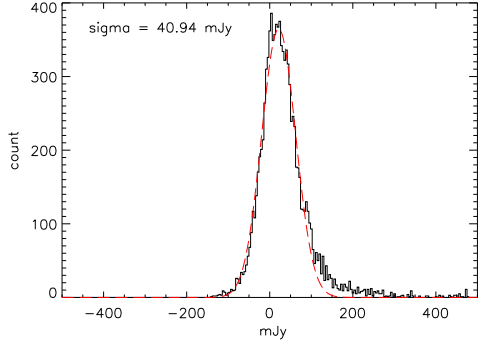} 
     \caption[]{\label{fig:confusion_measureMarano}10000 random
aperture photometry measurements on the FSM map indicating the confusion noise. The small
excess at high flux levels is due to real sources in the data.}
     \end{center}
\end{figure}

\begin{figure}
     \begin{center}
     \epsfxsize=7cm
     \epsfbox{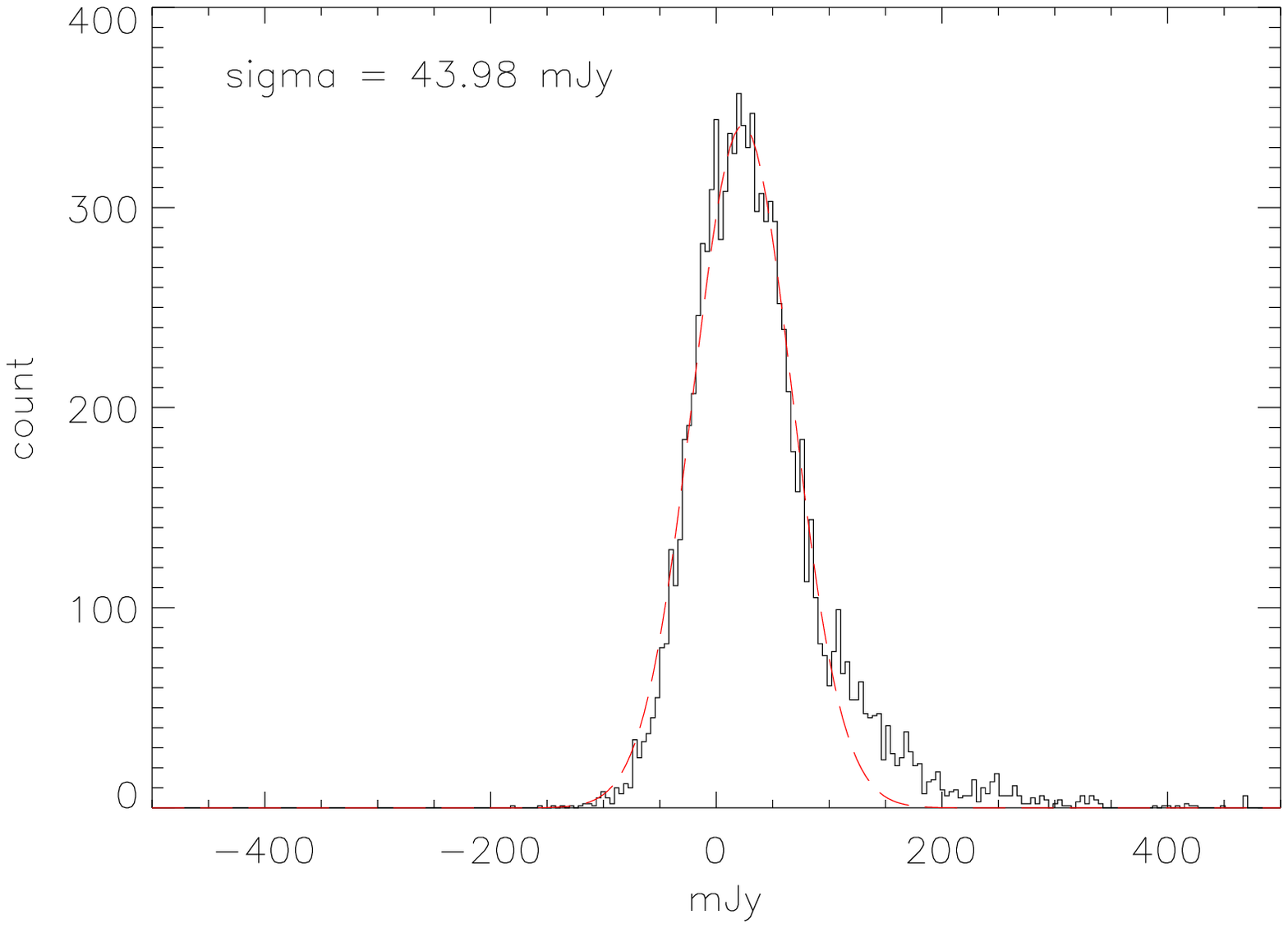}
     \caption[]{\label{fig:confusion_measureELAIS_N1}10000
random aperture photometry measurements on FN1.}
     \end{center}
\end{figure}

\begin{figure}
     \begin{center}
     \epsfxsize=7cm
     \epsfbox{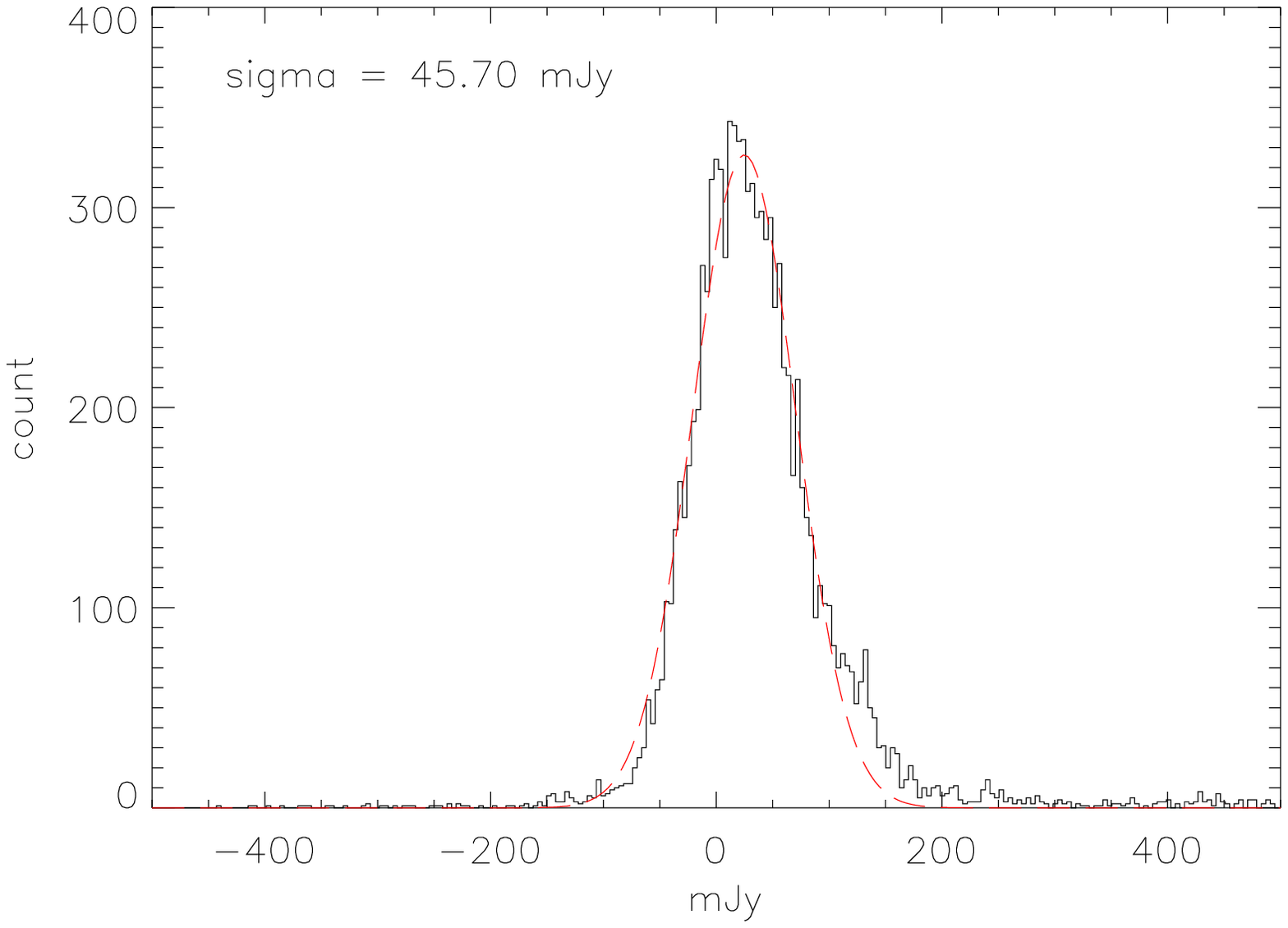} 
     \caption[]{\label{fig:confusion_measureELAIS_N2}10000
random aperture photometry measurements on FN2.}
     \end{center}
\end{figure}

We made 10000 measurements on each field at random positions, and obtained
distributions which are shown on figures
\ref{fig:confusion_measureMarano}, \ref{fig:confusion_measureELAIS_N1}
and \ref{fig:confusion_measureELAIS_N2}. These distributions
represent the probability of measurements by aperture photometry
on a field with sources and dominated by confusion.
They are fitted in their central part
by a gaussian, whose dispersion is an estimate of the confusion
noise. The distributions are plotted in
Fig.~\ref{fig:confusion_measureMarano} to
\ref{fig:confusion_measureELAIS_N2}. The assymetric part at high flux levels
reflects the counts of bright sources.
We finally derive $\sigma_c \simeq 45$ mJy for the confusion noise in all
of the \F fields (41 mJy for FSM, 44 for FN1 and 46 mJy for FN2).
The $3 \sigma_c$ level is thus 135 mJy and $4 \sigma_c$ 180 mJy.

%________________________________________________________________
% FIGURE RANDOM MEASUREMENTS
%________________________________________________________________
% FIGURE SIMULATION HISTO FLUX
\begin{figure}
     \begin{center}
     \epsfxsize=7cm 
     \epsfbox{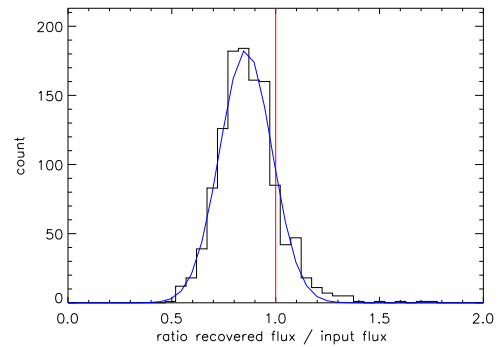} 
     \caption[]{\label{fig:analyse_sim_ELAIS_N10500mJy}Histogram
of the ratio of measured flux to input flux, when sources of 500 mJy are added to the maps.} 
     \end{center}
\end{figure}
This estimate is compatible with the classical definition of the
confusion, by computing the number of independent beams in all the \F
fields: with a FWHM of 94 arcseconds at 170 \mic in a $3.89$  sq. deg. 
surface, we have about $5700$ independent beams.
At the $3 \sigma$ limit, that is above 135 mJy, we have 196 sources (see
Sect.~\ref{catalog}), there are about 29 beams 
per source --- in good agreement with the classical definition of the confusion of
30 independent beams per source for sources brighter that $3
\sigma_c$. If we have a catalogue cutoff at $4 \sigma_c$ (resp. $5 \sigma_c$),
we obtain 54 (resp 91) independents beams per source. Our analysis is
compatible with the simulations of \cite{hogg2000}, who shows that 30
beams per source is a minimum where source counts are steep, and
suggests a threshold at about 50 beams per source. 

The cirrus fluctuations have a low probability of creating spurious sources at this
level of HI column-density, as shown in previous works, such as
\cite{gautier92}, \cite{lagache98a}, \cite{kawara98}, \cite{puget99},
and \cite{juvela2000}.
%
%________________________________________________________________
\subsection{Detector Noise}
The first field to be observed in our investigations was FSM1, and the goal was to
demonstrate the ability of doing a deep far infrared survey limited by
confusion rather than detector noise. With four independent
rasters mapping exactly the same sky, that is 16 independent
measurements, \cite{lagache98a} and \cite{puget99} show that the
detector noise level is about 3 mJy $1 \sigma$, i.e. far below the
confusion noise and thus neglected.
%________________________________________________________________
% FIGURE CATALOG SOURCES ON MAP
\begin{figure*} %[!h]
     \begin{center}
     \epsfxsize=13cm
     \epsfbox{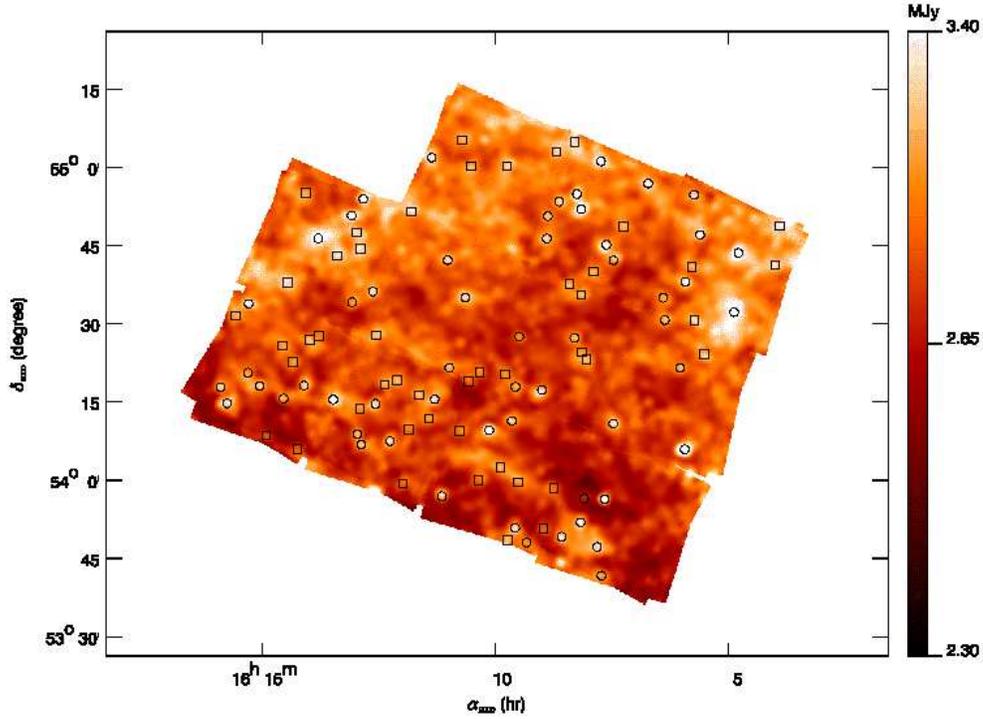} 
     \caption[]{\label{fig:proj_source_FN1_cifsc}Detected
sources on FN1 field. Circles are sources from the ISO FIRBACK Source
Catalog ($S_{\nu} > 180$ mJy) and squares are sources from the
Complementary ISO FIRBACK Source Catalog ($135 < S_{\nu} < 180$ mJy).}
     \end{center}
\end{figure*}
%
%________________________________________________________________
\subsection{Photometric Accuracy}
The histograms of the ratio of recovered flux to input flux of the simulated sources are used to
estimate the offset and the error
of the fluxes. One of these histograms is shown in figure
\ref{fig:analyse_sim_ELAIS_N10500mJy} for the FN1 field and 500 mJy
sources. 

One can see a systematic offset of the distribution's peak with
respect to the input flux.  This offset is constant for a given field,
and equals 16\%, 19\%, 18\% and 16\%
for the FN1, FN2, FSM1 and FSM234 fields, respectively.
The possible explanations for this offset are (1) the variation of the effective footprint inside the
field (due
to an inhomogeneous sampling of the sky) and (2) the loss of
flux at the edges of the pixels. 
We apply this correction on the source fluxes.

\begin{figure}
     \begin{center}
     \epsfxsize=7cm
     \epsfbox{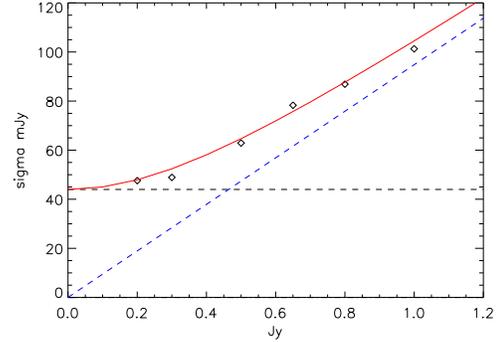}
     \caption[]{\label{fig:sigma_flux_sources}Evolution of
$\sigma_s$, standard deviation of measured flux
on the histograms, as a function of the source flux
(diamonds). $\sigma_s$ can be decomposed in two components: (1) a
constant component due to confusion noise $\sigma_c$ (horizontal dashed line)
and (2) a component proportional to the flux $\sigma_p$
(sloped dashed line). $\sigma_s$ is fitted by $\sqrt{\sigma_c^2 + \sigma_p^2}$
(solid line).}
     \end{center}
\end{figure}

The standard deviation of the fitted gaussian, $\sigma_s$, estimates the
dispersion of the source flux measurements. Figure
\ref{fig:sigma_flux_sources} shows the variation of $\sigma_s$ in mJy
as a function of the source flux in Jy, in the FN1 field; the
variation is similar in the other fields. $\sigma_s$ can be decomposed in two components:
\begin{itemize}
\item[$\bullet$] a constant component due to confusion noise $\sigma_c$ 
\item[$\bullet$] a component ($\sigma_p$) proportional to the source flux, due to
the difference between the mean effective footprint and the local
effective footprint.
\end{itemize}
The data points are fitted by the quadratic sum of the constant and
the proportional component $\sqrt{\sigma_c^2 + \sigma_p^2}$.

The source flux uncertainties are computed for each field;
however, there is little field-to-field variation.
The uncertainty in the source flux is about 25\%
near $3\sigma_c$ at low fluxes, about 20\%
near $5\sigma_c$ and decreases to about 10\%
at high flux levels(near 1 Jy). 

%________________________________________________________________
% FIGURE SIGMA SOURCES
%________________________________________________________________
% FIGURES SIMULATION HISTO DISTANCES
\begin{figure}
     \begin{center}
     \epsfxsize=7cm 
     \epsfbox{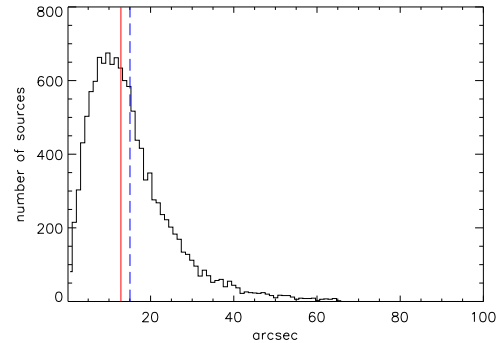}
     \caption[]{\label{fig:dist_sim_all}Histogram of distances of
identifications in the simulations. All sources brighter than 500 mJy (where the sample is
complete) in the three \F \, fields are shown. The solid line corresponds to the median at 13
arcseconds and the dashed line at 15 arcseconds.}
     \end{center}
\end{figure}
%
%________________________________________________________________
\subsection{Positional Accuracy}
The identification of the sources in the simulations allows us to
derive the positional accuracy. 
We neglect the telescope absolute pointing error of 1" \cite[]{kessler2000}.
Fig.~\ref{fig:dist_sim_all} shows
the distribution of the distance offset between the input source and
the extracted source positions.
All sources brighter than 500 mJy --- i.e. where the sample is complete
(see Sect.~\ref{completeness}) --- are recovered inside a 65"
radius: the mean recovered distance is 15", and 90\%
of the sample falls inside 28". Taking all the sources with
flux levels brighter than 180 mJy, 90\% of the sample is recovered inside a radius of 42". 
We conclude that 99\% (respectively 93\%) of the sources are found in a circle
of radius of 50", and 98\% (respectively 90 \%) in 42"
when the sample is complete, above 500 mJy (respectively 180 mJy).

%
%________________________________________________________________
\section{\F Source Catalogs}
\label{catalog}
%
%________________________________________________________________
\subsection{ISO \F Source Catalog}
The final catalog, the {\it ISO \F Source Catalog} (IFSC), contains
106 sources with fluxes between 180 mJy ($4 \sigma$) and 2.4 Jy.
The catalog is given for each field in tables \ref{catalog_marano} 
to \ref{catalog_elaisn2}. All the sources have been checked for
detection in all individual measurements.
It is interesting to note that above $5 \sigma_c$ the source density
is constant in the fields, with 16 sources in FSM, 15 in FN2, and 32
sources in FN1 which is twice the size of the other fields. The source
density is thus $16 \pm 4$ sources brighter than 225 mJy per square
degree. At the $4 \sigma_c$ limit, the source density is $27 \pm 5$
sources brighter than 180 mJy per square
degree, with a larger field-to-field dispersion. 
The brightest sources in FSM lie at 497 and 443 mJy, in FN1 at 838,
597 and 545 mJy, and in FN2 at 2377, 1251, 803, 682, 666 and 522 mJy.
%________________________________________________________________
% FIGURE: COMPLETENESS
\begin{figure}
     \begin{center}
     \epsfxsize=7cm 
     \epsfbox{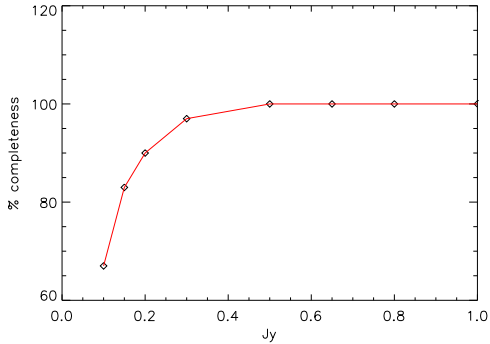} 
     \caption[]{\label{fig:completeness_counts}Completeness of the \F
catalog, computed from the simulations as the ratio of the number of detected sources to the
number of added sources.}
     \end{center}
\end{figure}
%________________________________________________________________
% FIGURE: EDDINGTON-MALMQUIST
\begin{figure}
     \begin{center}
     \epsfxsize=7cm 
     \epsfbox{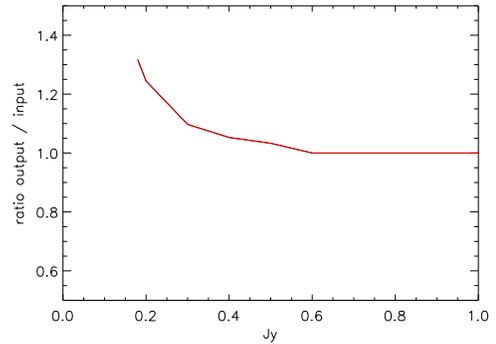}  
     \caption[]{\label{fig:eddington_malmquist_counts}Malmquist-Eddington
bias. Ratio of simulated source counts to simulated observed source counts. Due to flux
uncertainties, the number of counts is overestimated at low flux levels.}
     \end{center}
\end{figure}
%
%________________________________________________________________
\subsection{Complementary ISO \F Source Catalog}
Sources with flux levels above the $3\sigma_c$ limit are 
higher redshift candidates and they can be used for statistical study of the nature of 170 \mic\,
sources. Nevertheless, the lower signal to confusion-noise ratio leads to lower flux accuracy ---
reduced to about 25\%
at 135 mJy --- and may include spurious sources: this larger uncertainty suggests avoiding the use
of these sources, e.g. in the counts.

Candidates for $z > 1$ may be selected on the basis of photometric
redshift using the FIR-radio correlation \cite[]{condon92,helou85} and
the submillimetre-radio correlation \cite[]{carilli2000a}. The success of
recent submillimetre detections of \F sources with SCUBA at the JCMT
\cite[with an rms sensitivity of 2 mJy, ]{scott2000} and with MAMBO at
IRAM-30m (with an rms sensitivity better than 0.5 mJy, Lagache et al.,
in prep) in the millimetre range confirms the relevance of this technique.

% Input des fichiers tex crees par plot_lognlogs.pro
 
% file /home/hdole/DATA/ISO/firback/catalog/nbsources_lognlogs_aa.tex
 
\begin{table}[!ht]
        \caption{Number of sources per flux bin in $3.89^{\circ 2}$ used for the source counts, without any correction.}
        \label{nbsources_lognlogs}
        \begin{center}
        \leavevmode
        \begin{tabular}[h]{cccc}
                \hline \\[-5pt]
                flux min  & flux max & number per      &  cumulative \\
                 (mJy)      &   (mJy)    & bin             &  number \\[+5pt]
                \hline \\[-5pt]
 180.0 &  190.0 &  13 &  106 \\
 190.0 &  210.0 &  20 &   93 \\
 210.0 &  240.0 &  21 &   73 \\
 240.0 &  300.0 &  24 &   52 \\
 300.0 &  500.0 &  19 &   28 \\
 500.0 & $ \infty $ &   9 &    9 \\
        \hline \\[-5pt]
        \end{tabular}
        \end{center}
\end{table}

% created with plot_lognlogs.pro by Herve Dole on Thu Jan  4 15:19:52 2001

%
In this frame of mind, we compile a Complementary ISO \F Source Catalog
(CIFSC, tables \ref{compl_catalog_marano} to
\ref{compl_catalog_elaisn2} for each field)
 which contains 90 sources whose flux levels lie in the range 135 to
180 mJy (3 to $4\sigma_c$). 
All the sources have been checked for
detection in all individual measurements.
There are 15 sources in FSM, 47 in FN1, and 28 in FN2. 
As an example, Fig~\ref{fig:proj_source_FN1_cifsc} shows of the detected
sources in the FN1 field.
At this flux
level, the source density is not constant between the fields and 
fluctuates at about $23 \pm 8$ sources per square degree in the 
range 135 - 180 mJy. 
%
%________________________________________________________________
\section{Source Counts}
\label{counts}
%
%________________________________________________________________
\subsection{Completeness}
\label{completeness}
Simulations allow us to derive the completeness, that is the ratio at
a given flux between the number of added sources and the number of detected ones. The
completeness is plotted in Fig.~\ref{fig:completeness_counts}. Our
sample is complete above 500 mJy, and is about 90\% (respectively about 85\%)
complete above 225 mJy (respectively 180 mJy). We thus correct the surface source density for
this incompleteness.
%
%________________________________________________________________
\subsection{Malmquist-Eddington Bias}
\label{malmquist}
Uncertainties in the flux determination introduce an excess in the
number of counts, known as the Malquist-Eddington bias. 
We characterize it with the results of the simulations, by comparing the effect of a
flux dispersion on a known input source count
model: a simple power law. Fig.~\ref{fig:eddington_malmquist_counts} shows the ratio
of an input source count model, to the simulated observations
of this model. We apply the appropriate correction to the data:
at 225 mJy (respectively 180 mJy) the raw counts have to be decreased by
20\% (respectively 30\%).
We check that these values are not more sensitive than 5\% (respectively 10\%)
at $5 \sigma_c$ (respectively $4 \sigma_c$) to the power
law of the input model in the range 3.0 - 3.6.
%
%________________________________________________________________
\subsection{\F Source Counts}
%
%________________________________________________________________
% FIGURE FIRBACK COUNTS DATA ALONE
\begin{figure*}[!ht]
     \begin{center}
     \epsfxsize=13cm 
     \epsfbox{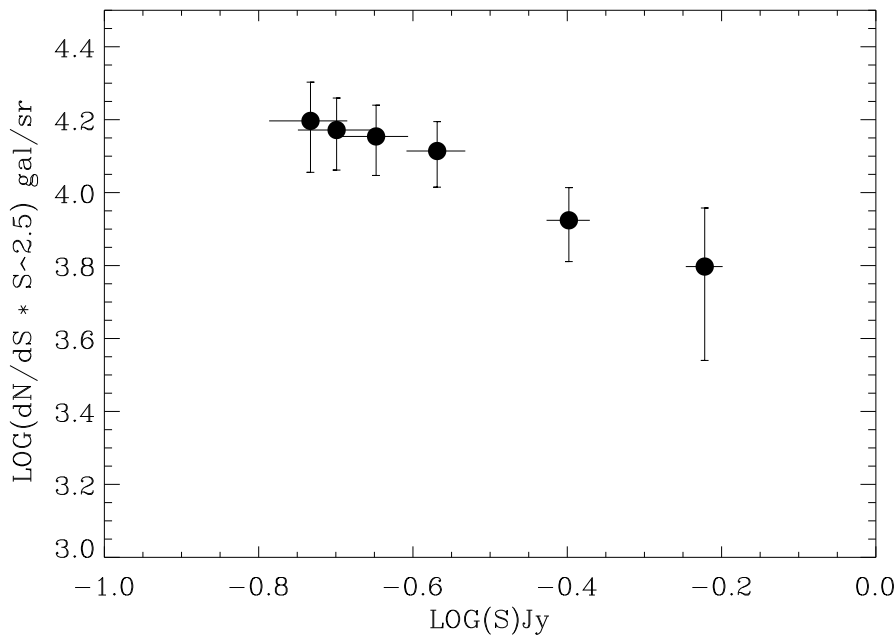} 
     \caption[]{
\label{fig:counts_alone_diff}\F differential source counts (normalized to Euclidian counts) at 170
\mic. 106 sources are brighter than 180 mJy ($4 \sigma_c$) on
$3.89$ sq. deg. The slope of the differential source counts is $3.3 \pm 0.6$ between 180 and
500 mJy.}
     \end{center}
\end{figure*}
Figure~\ref{fig:counts_alone_diff} shows the differential source counts at 170\mic\, coming from
the \F survey ($3.89$ sq. deg.), with 106 sources between 180 ($4
\sigma$) and 2400 mJy. The 
horizontal error bar gives the flux uncertainty, and the vertical
error bar the poisson noise in $\sqrt{n}$ where $n$ is the number
of sources in the bin.

The statistics of sources used for source counts before any correction is given in
Table~\ref{nbsources_lognlogs}. The integral (respectively differential) source count values
are given in Table~\ref{plot_lognlogs} (respectively Table~\ref{plot_lognlogsdiff}). Note that
for the differential
counts we took only 5 sources in the last flux box, corresponding to
highest fluxes (between 500 and 700 mJy). 

The two points at high flux levels are compatible with no evolution since we
can adjust a horizontal line inside the error bars.
The slope of the differential source counts is not constant, but can
reasonably be fitted by a linear of slope $3.3 \pm 0.6$ 
between 180 and 500 mJy.
% Input des fichiers tex crees par plot_lognlogs.pro
 
% file /home/hdole/DATA/ISO/firback/catalog/plot_lognlogs_aa.tex
 
\begin{table}[!ht]
        \caption{FIRBACK Integrated Source Counts.}
        \label{plot_lognlogs}
        \begin{center}
        \leavevmode
        \begin{tabular}[h]{cc}
                \hline \\[-5pt]
                $log_{10}$ of galaxy density   & flux  \\
                $(sr^{-1})$                      & (mJy)      \\[+5pt]
                \hline \\[-5pt]
  $4.919 \pm 0.085 $ & $    180.0 \pm  21.5  $ \\      
  $4.869 \pm 0.090 $ & $    190.0 \pm  21.8  $ \\      
  $4.774 \pm 0.102 $ & $    210.0 \pm  22.6  $ \\      
  $4.638 \pm 0.121 $ & $    240.0 \pm  23.6  $ \\      
  $4.374 \pm 0.166 $ & $    300.0 \pm  25.8  $ \\      
  $3.894 \pm 0.301 $ & $    500.0 \pm  33.0  $ \\      
        \hline \\[-5pt]
        \end{tabular}
        \end{center}
\end{table}

% created with plot_lognlogs.pro by Herve Dole on Thu Jan  4 15:19:52 2001

%
%
%________________________________________________________________
\section{Discussion}
%
%________________________________________________________________
\subsection{Comparison with other Work}
\cite{kawara98} estimated the confusion level to be 45 mJy, and extracted
45 sources brighter than 150 mJy ($3 \sigma_c$) in the $1.1$ sq. deg. 
Lockman Hole field. \cite{juvela2000} found $\sigma_c = 44$
mJy, and detected 55 sources brighter than 150 mJy in $1.5$ sq. deg.
Both these estimates are consistent with our measurements.

Our raw results are in agreement with the pioneering work on
$1/16^{th}$  of the area of the entire \F survey by
\cite{lagache98a} and \cite{puget99}. Without completeness or
Malmquist-Eddington bias correction, our catalogs are similar.
Of the 24 sources of \cite{puget99}, we detect 18. The six
missing sources are: (1) on the edges of the field with fewer
observations than required in our procedure of extraction for three of
their sources, and (2) in more confused regions for the other three.

For the 18 common sources, the photometry is in excellent agreement
(except for one source which is near the edge of the field). 
Both analyses find 13 sources at fluxes higher than 150 mJy in this field.

In our preliminary work \cite[]{dole99,dole2000}, we detected the
sources by eye and used the same photometry as \cite{puget99} for
consistency, but we did not remove bright sources to measure the
fainter sources. Statistically, these efforts are compatible with our
current source counts.
%
%________________________________________________________________
\subsection{Comparison with Models}

% Input des fichiers tex crees par plot_lognlogs.pro
 
% file /home/hdole/DATA/ISO/firback/catalog/plot_lognlogsdiff_aa.tex
 
\begin{table}[!ht]
        \caption{FIRBACK Differential Source Counts.}
        \label{plot_lognlogsdiff}
        \begin{center}
        \leavevmode
        \begin{tabular}[h]{cc}
                \hline \\[-5pt]
                $log_{10}$ of $\frac{dN}{dS} \times S^{2.5}$    & flux bin \\
                $ \left ( sr^{-1} \times Jy^{1.5} \right ) $   & (mJy)      \\[+5pt]
                \hline \\[-5pt]
  $4.179 \pm 0.247 $ & $   180 - 190  $ \\      
  $4.157 \pm 0.198 $ & $   190 - 210  $ \\      
  $4.143 \pm 0.193 $ & $   210 - 240  $ \\      
  $4.107 \pm 0.180 $ & $   240 - 300  $ \\      
  $3.921 \pm 0.203 $ & $   300 - 500  $ \\      
  $3.797 \pm 0.418 $ & $   500 - 700  $ \\      
        \hline \\[-5pt]
        \end{tabular}
        \end{center}
\end{table}

% created with plot_lognlogs.pro by Herve Dole on Thu Jan  4 15:19:52 2001

\begin{figure}[!ht]
     \begin{center}
     \epsfxsize=8cm
     \epsfbox{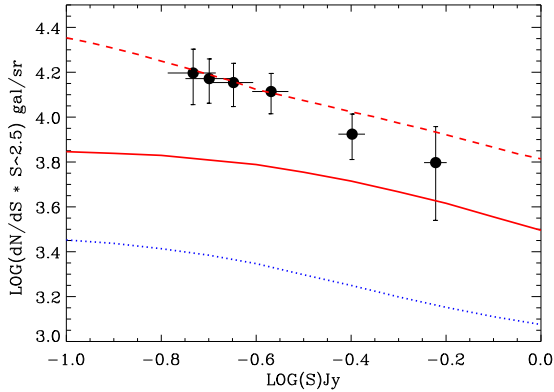} 
     \caption[]{\label{fig:counts_models3_diff}\F differential source
counts at 170 \mic\, with models from \cite{guiderdoni98} with evolution and without ULIRG's
(A, dotted line) and with evolution with ULIRG's (E, solid line), and from \cite{dole2000} 
(strongly evolving LIRGs, dashed line).} 
     \end{center}
\end{figure}

\begin{figure}[!ht]
     \begin{center}
     \epsfxsize=8cm
     \epsfbox{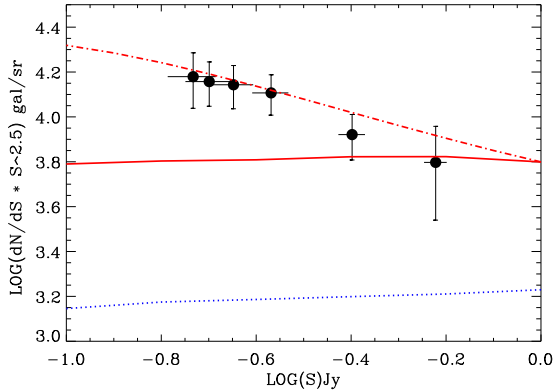} 
     \caption[]{\label{fig:counts_models4_diff}\F differential source
counts at 170 \mic\, with models from \cite{franceschini98} without evolution (dotted line) and
with evolution (solid line), and from \cite{rowan-robinson2001}
(pure luminosity evolution, dot-dashed line).} 
     \end{center}
\end{figure}

%
The semi-analytical model from \cite{guiderdoni98} was used in the \F proposal to
justify the integration time and surface coverage, and has been
improved recently \cite[]{devriendt2000}. Our phenomenological model \cite[]{dole2000}
was developed by taking into account all the observational constraints in the
infrared and submillimetre range, and is based on strong evolution
of a bright population of galaxies. Both models are presented in
Fig.~\ref{fig:counts_models3_diff}. The models of \cite{franceschini98}, with and without
evolution, are shown in Fig.~\ref{fig:counts_models4_diff} together
with the pure luminosity evolution model of \cite{rowan-robinson2001}.

The data unambiguously reject models without evolution or with
low evolution at flux levels fainter than 500 mJy. The no-evolution model of
\cite{franceschini98} (dots in Fig.~\ref{fig:counts_models4_diff}) and
the model without ULIRGs \cite[]{sanders96} of \cite{guiderdoni98} (dotted line in
Fig.~\ref{fig:counts_models3_diff}) are incompatible with the data:
they predict between 5
and 10 times fewer sources than observed.

%________________________________________________________________
% FIGURE FIRBACK COUNTS WITH MODELS
%________________________________________________________________
% FIGURE FIRBACK COUNTS WITH MODELS

Model E of \cite{guiderdoni98} with
strong evolution and an addition of ULIRGs underestimates the source
counts by a factor of 2, and predicts a lower slope than the
observations. Nevertheless, the agreement within a factor
of 2 between model E and the final observed source counts is quite
remarkable: this model was developed to account for the CIB, and was used
for predicting the \F source counts at the time of submission of this observing program.
The phenomenological model of \cite{dole2000} fits the data at faint
fluxes, as well as the model of \cite{rowan-robinson2001}.

Other semi-analytical models like e.g. \cite{blain99c},
or phenomenological models like e.g. those of \cite{tan99}, \cite{xu2000},
and \cite{pearson2000}, also try to reproduce the
spectrum of the CIB as well as the source counts in the whole spectral
domain from the mid infrared (sometimes optical) to the sub-millimeter
(sometimes centimeter) range. It is beyond the scope of this paper to
compare all these models with our observations, but scenarios without
strong evolving populations of LIRGs are uniformly unable to reproduce the data.
%
%________________________________________________________________
\subsection{Resolving the Cosmic Infrared Background at 170~\mic}
We now ask what fraction of the CIB is contributed by sources brighter than 135 mJy at 170 \mic
~?
Since the flux integral is dominated by sources at lower flux levels, it is rather
simple to compute its value on the assumption that source
counts have a constant slope.
We estimate that $4 \pm 1$\% of the CIB is resolved
into sources brighter than 135 mJy.

Using the model of \cite{dole2000}, we show that 7\%
of the CIB is resolved in sources brighter
than 135 mJy.

Thus, the population of individually observed sources in the \F survey does not dominate the
CIB at this wavelength.\\

We can also ask at which flux levels the CIB will be largely resolved at 170 \mic \,? 
The observed slope of the source counts may be extrapolated to lower
flux levels to predict a convergence. The expected flattening of the source counts
close to the convergence is neglected, and thus the derived values
give an upper limit. With this proviso, we find that the 170 \mic \, background should be resolved
at flux levels in the range 10 to
20 mJy, an order of magnitude fainter than the ISO
sensitivity. 

Using our model presented in Fig.~\ref{fig:counts_models3_diff},
we predict that 80 to 90\% of the CIB should be resolved in the range 2 to 5
mJy.

The required sensitivity at 170 \mic, about 60 times better than ISO, is not
reachable with the 1m-class space infrared observatories such as
NASA's SIRTF. SIRTF should be able to break about 15-30\% of the
background into discrete sources at 160 \mic.
4m-class ESA's Herschel (former FIRST) should be able to break the
bulk of the background.

%
%________________________________________________________________
\subsection{Survey Optimisation and Confusion}
While the confusion is found to be identical in the 3 \F fields at about 45 mJy (regardless of
sampling), the FSM fields have been observed twice as frequently as the northern fields. This
means that the confusion limit was
reached faster than expected. Two years after the end of the
observations, and five years after the launch of ISO, our analysis shows
that the best (ideal) observational strategy would have been to 
repeat the individual observations (rasters) 4 times to obtain enough redundancy (done in
the FSM field) with less integration and proper oversampling. 
With this optimisation, we could have gained 25\% more surface with
the earned time, or performed complementary observations at another wavelength.

An early proper determination of the confusion level is thus a
key factor for extragalactic infrared surveys from space, since the
confusion is high due to the strong evolution, and
limits the surveys.
It is challenging, given the relatively short time spend in the 
``Performance Verification'' or ``In Orbit Checkout'' phases that
normally predece routine astronomical observations in space.
%
%______________________________________________________________
\section{Conclusion \& Summary}
The analysis of the \F ISO deep survey sources at 170 \mic \, is presented.
After a process of data reduction and calibration of extended emission
\cite[]{lagache2001}, we performed extensive simulations to validate
our source extraction process, and studied the sources of
noise and accuracy in photometry and astrometry. The confusion
$\sigma_c$ equals 45 mJy.

We compiled the ISO \F Source Catalog ($S_{170} > 4 \sigma_c$) and the
Complementary ISO \F Source Catalog ($3 \sigma_c < S_{170} < 4
\sigma_c$, for follow-up purposes) containing 196 sources.
It is important to note that the extended source calibration is in excellent agreement
with DIRBE and the point source calibration is in agreement with IRAS.
The differential source counts show a steep slope of $3.3 \pm 0.6$
between 180 and 500 mJy, and a significant excess of
faint sources with respect to low or moderate evolution expectations. 

The steep slope of the source counts has important consequences on
the sensitivity limits of the deep surveys conducted in the far
infrared: the confusion noise is large, as it will be for future
observatories, and will impact dramatically on the future IR deep surveys.

%________________________________________________________________
%\subsection{Nature of the \F Sources}
One important intention of the \F survey was to probe the nature of the extragalactic far-infrared
sources.
According to most of the models, the steep slope of the source counts is due
to a strongly evolving population of LIRGs. Our model shows that the
effect of the K-correction alone is insufficient to explain the
observations.  To definitively investigate this question, one has to identify the sources
and understand their nature.
Discussions of the nature of the \F sources is beyond of the scope of this
paper and will be discussed elsewhere. The multiwavelength follow-up performed at 1.4 GHz, 1.3 mm,
850 \& 450 \mic, as well as other ISO and optical / NIR data, seems to
show that most of the sources (typically 50\%) are local
($z < 0.3$), and about 10\% at high redshift ($z > 1$). Massive star formation
seems also to be dominant. Nevertheless, identifying \F sources
is not easy because of the uncertainty in the positions at 170 \mic. \\

The summary of the \F survey is as follows:
%
%______________________________________________________________
\begin{itemize}
\item[$\bullet$] observation of about $4$ sq. deg. in 3 high galactic latitude fields: FSM, FN1
\& FN2
\item[$\bullet$] ISOPHOT AOT P22 raster map mode with the \verb|C_200| array and the
\verb|C_160| filter at 170 \mic
\item[$\bullet$] 128 or 256 seconds of integration per sky pixel
\item[$\bullet$] extraction of instrumental effects: long and short term transients, photometric
correction
\item[$\bullet$] calibration of extended emission: excellent agreement between PHT and
DIRBE
\item[$\bullet$] calibration of point sources compatible with IRAS
\item[$\bullet$] instrumental noise: 3 mJy $1 \sigma$
\item[$\bullet$] confusion noise: 45 mJy $1 \sigma$; $4 \sigma_c$ sensitivity: 180 mJy
\item[$\bullet$] ISO \F Source Catalog: 106 sources between 180 mJy and 2.4 Jy
\item[$\bullet$] Complementary ISO \F Source Catalog: 90 sources between 135 and 180 mJy
\item[$\bullet$] Flux uncertainty error: 25\% at $3\sigma_c$, 20\% at $5 \sigma_c$, and reduced
to 10\% at higher flux levels
\item[$\bullet$] positional error: 100 arcsecond diameter circle (99\% of the sources)
\item[$\bullet$] source density for $S_{170} > 225$ mJy: $16 \pm 4$ sources per square degree
\item[$\bullet$] source density for $S_{170} > 180$ mJy: $27 \pm 5$ sources per square degree
%\item[$\bullet$] source density for $S_{170} > 135$ mJy: $50 \pm 10$ sources per square
degree
\item[$\bullet$] slope of the differential source counts: $3.3 \pm 0.6$ between 180 and 500 mJy
\item[$\bullet$] 4 to 7\% of the Cosmic Infrared Background at 170
\mic\, is resolved into sources brighter than 135 mJy
\item[$\bullet$] Prediction that the CIB will be resolved at flux levels in the range 1 to 10 mJy at
170 \mic
\item[$\bullet$] Catalogs, images, and plots available on line
at: \verb+http://wwwfirback.ias.u-psud.fr+ 
\end{itemize}

%
%______________________________________________________________
\begin{acknowledgements}
HD, GL \& JLP appreciate discussions with Rene Laureijs and 
Carlos Gabriel at Vilspa and Ulrich Klaas at
Heidelberg. We also are greatful
to Alain Abergel, Alain Coulais \& Marc-Antoine Miville-Desch\^enes at
IAS, for stimulating and helpful discussions through the
analysis of the \F data. Thanks also go to Martin Kessler and his team
who did a great job in planning ISO observations so efficiently. MH acknowledges support of his
participation on ISO through NASA grants and contracts. DE acknowledges support from NASA grant NAG5-8218.

\end{acknowledgements}

%
%________________________________________________________________
% TABLES FINAL CATALOG
% Input des fichiers tex crees par sort_pht_catalog.pro
%------------------------------------------------------
% IFSC /home/hdole/DATA/ISO/firback/catalog/Marano/cat2count_marano_aa.tex
 
\begin{table} [!ht]
        \caption{\F\, Catalog in FSM: coordinates are in hours ($\alpha_{2000}$) or degrees ($\delta_{2000}$), minutes, seconds, the flux $S$ and the flux uncertainty $\delta S$ at $170 \mu m $ are in mJy.}
        \label{catalog_marano}
        \begin{center}
        \leavevmode
        \begin{tabular}[h]{cccccc}
                \hline \\[-5pt]
                source       & $\alpha_{2000}$    & $\delta_{2000}$  & $S$  & $\delta S$ & $S/\sigma_c $\\[+5pt]
                \hline \\[-5pt]
FSM\_000 &   3 09 25 & -54 52 04 &   497 &   52 &  11.0 \\
FSM\_001 &   3 12 07 & -55 17 09 &   443 &   50 &   9.9 \\
FSM\_002 &   3 12 29 & -55 16 30 &   420 &   49 &   9.3 \\
FSM\_003 &   3 11 59 & -55 14 20 &   369 &   46 &   8.2 \\
FSM\_004 &   3 08 37 & -55 20 45 &   365 &   46 &   8.1 \\
FSM\_005 &   3 10 22 & -54 31 55 &   301 &   43 &   6.7 \\
FSM\_006 &   3 10 45 & -54 32 05 &   300 &   43 &   6.7 \\
FSM\_007 &   3 12 10 & -55 09 00 &   296 &   43 &   6.6 \\
FSM\_008 &   3 12 33 & -54 57 00 &   269 &   42 &   6.0 \\
FSM\_009 &   3 08 42 & -54 27 28 &   267 &   41 &   5.9 \\
FSM\_010 &   3 10 16 & -55 01 37 &   261 &   41 &   5.8 \\
FSM\_011 &   3 12 53 & -55 09 28 &   239 &   40 &   5.3 \\
FSM\_012 &   3 08 03 & -54 34 33 &   232 &   40 &   5.2 \\
FSM\_013 &   3 15 18 & -55 01 26 &   228 &   40 &   5.1 \\
FSM\_014 &   3 14 50 & -54 59 09 &   226 &   40 &   5.0 \\
FSM\_015 &   3 10 37 & -54 26 16 &   225 &   40 &   5.0 \\
FSM\_016 &   3 13 07 & -54 49 40 &   214 &   39 &   4.8 \\
FSM\_017 &   3 08 24 & -54 28 04 &   210 &   39 &   4.7 \\
FSM\_018 &   3 10 01 & -55 11 45 &   207 &   39 &   4.6 \\
FSM\_019 &   3 07 28 & -55 09 07 &   202 &   39 &   4.5 \\
FSM\_020 &   3 09 31 & -55 25 04 &   200 &   38 &   4.4 \\
FSM\_021 &   3 08 50 & -55 05 45 &   190 &   38 &   4.2 \\
FSM\_022 &   3 09 24 & -55 10 37 &   182 &   38 &   4.1 \\
        \hline \\[-5pt]
        \end{tabular}
        \end{center}
\end{table}

% created with sort_pht_catalog.pro by Herve Dole on Fri Dec 29 16:32:13 2000

% IFSC /home/hdole/DATA/ISO/firback/catalog/ELAIS_N1/cat2count_elaisn1_aa.tex
 
\begin{table} [!ht]
        \caption{\F\, Catalog in FN1: coordinates are in hours ($\alpha_{2000}$) or degrees ($\delta_{2000}$), minutes, seconds, the flux $S$ and the flux uncertainty $\delta S$ at $170 \mu m $ are in mJy.}
        \label{catalog_elaisn1}
        \begin{center}
        \leavevmode
        \begin{tabular}[h]{cccccc}
                \hline \\[-5pt]
                source       & $\alpha_{2000}$    & $\delta_{2000}$  & $S$  & $\delta S$ & $S/\sigma_c $\\[+5pt]
                \hline \\[-5pt]
FN1\_000 &  16 05 52 &  54 06 46 &   838 &   90 &  18.6 \\
FN1\_001 &  16 07 37 &  53 57 25 &   597 &   73 &  13.3 \\
FN1\_002 &  16 10 07 &  54 10 40 &   545 &   69 &  12.1 \\
FN1\_003 &  16 12 55 &  54 54 57 &   408 &   59 &   9.1 \\
FN1\_004 &  16 11 09 &  53 58 01 &   391 &   58 &   8.7 \\
FN1\_005 &  16 04 44 &  54 32 56 &   374 &   57 &   8.3 \\
FN1\_006 &  16 04 37 &  54 44 16 &   348 &   55 &   7.7 \\
FN1\_007 &  16 13 32 &  54 16 22 &   338 &   54 &   7.5 \\
FN1\_008 &  16 08 58 &  54 18 25 &   335 &   54 &   7.4 \\
FN1\_009 &  16 08 05 &  54 53 02 &   313 &   52 &   7.0 \\
FN1\_010 &  16 09 34 &  53 51 57 &   309 &   52 &   6.9 \\
FN1\_011 &  16 08 09 &  53 52 58 &   304 &   52 &   6.8 \\
FN1\_012 &  16 12 17 &  54 08 31 &   302 &   51 &   6.7 \\
FN1\_013 &  16 07 38 &  55 02 13 &   300 &   51 &   6.7 \\
FN1\_014 &  16 15 51 &  54 15 18 &   295 &   51 &   6.6 \\
FN1\_015 &  16 07 25 &  54 11 52 &   294 &   51 &   6.5 \\
FN1\_016 &  16 07 32 &  54 46 12 &   289 &   51 &   6.4 \\
FN1\_017 &  16 05 48 &  54 38 56 &   288 &   50 &   6.4 \\
FN1\_018 &  16 14 11 &  54 19 01 &   288 &   50 &   6.4 \\
FN1\_019 &  16 12 36 &  54 15 39 &   285 &   50 &   6.3 \\
FN1\_020 &  16 08 11 &  54 55 58 &   283 &   50 &   6.3 \\
FN1\_021 &  16 13 11 &  54 51 43 &   271 &   49 &   6.0 \\
FN1\_022 &  16 16 00 &  54 18 25 &   270 &   49 &   6.0 \\
FN1\_023 &  16 08 33 &  53 50 16 &   270 &   49 &   6.0 \\
FN1\_024 &  16 09 38 &  54 12 28 &   266 &   49 &   5.9 \\
FN1\_025 &  16 08 35 &  54 54 32 &   243 &   47 &   5.4 \\
FN1\_026 &  16 14 37 &  54 16 26 &   241 &   47 &   5.3 \\
FN1\_027 &  16 11 25 &  55 02 59 &   234 &   47 &   5.2 \\
FN1\_028 &  16 07 42 &  53 42 43 &   229 &   46 &   5.1 \\
FN1\_029 &  16 11 19 &  54 16 37 &   229 &   46 &   5.1 \\
FN1\_030 &  16 05 28 &  54 47 52 &   228 &   46 &   5.1 \\
FN1\_031 &  16 11 03 &  54 43 19 &   225 &   46 &   5.0 \\
FN1\_032 &  16 12 41 &  54 37 11 &   224 &   46 &   5.0 \\
FN1\_033 &  16 13 00 &  54 09 50 &   224 &   46 &   5.0 \\
FN1\_034 &  16 07 23 &  54 43 12 &   221 &   46 &   4.9 \\
FN1\_035 &  16 15 25 &  54 34 30 &   218 &   45 &   4.8 \\
FN1\_036 &  16 06 36 &  54 57 54 &   214 &   45 &   4.7 \\
FN1\_037 &  16 15 09 &  54 18 46 &   210 &   45 &   4.7 \\
FN1\_038 &  16 07 48 &  53 48 14 &   207 &   45 &   4.6 \\
FN1\_039 &  16 08 50 &  54 51 46 &   205 &   45 &   4.6 \\
FN1\_040 &  16 09 28 &  54 28 40 &   205 &   45 &   4.5 \\
FN1\_041 &  16 08 15 &  54 28 22 &   204 &   44 &   4.5 \\
FN1\_042 &  16 10 39 &  54 36 10 &   202 &   44 &   4.5 \\
FN1\_043 &  16 05 57 &  54 22 26 &   201 &   44 &   4.5 \\
FN1\_044 &  16 09 33 &  54 19 01 &   198 &   44 &   4.4 \\
FN1\_045 &  16 08 51 &  54 47 27 &   198 &   44 &   4.4 \\
FN1\_046 &  16 12 55 &  54 07 48 &   196 &   44 &   4.4 \\
FN1\_047 &  16 08 04 &  53 57 32 &   196 &   44 &   4.4 \\
FN1\_048 &  16 11 00 &  54 22 40 &   192 &   44 &   4.3 \\
FN1\_049 &  16 13 09 &  54 35 05 &   186 &   43 &   4.1 \\
        \hline \\[-5pt]
        \end{tabular}
        \end{center}
\end{table}
\begin{table} [!ht]
        \caption{\F\, Catalog in FN1 (continued).}
        \label{catalog_elaisn1_1}
        \begin{center}
        \leavevmode
        \begin{tabular}[h]{cccccc}
                \hline \\[-5pt]
                source       & $\alpha_{2000}$    & $\delta_{2000}$  & $S$  & $\delta S$ & $S/\sigma_c $\\[+5pt]
                \hline \\[-5pt]
FN1\_050 &  16 13 55 &  54 47 16 &   185 &   43 &   4.1 \\
FN1\_051 &  16 05 35 &  54 55 37 &   185 &   43 &   4.1 \\
FN1\_052 &  16 06 16 &  54 31 37 &   183 &   43 &   4.1 \\
FN1\_053 &  16 09 19 &  53 49 08 &   182 &   43 &   4.0 \\
FN1\_054 &  16 15 25 &  54 21 17 &   182 &   43 &   4.0 \\
FN1\_055 &  16 06 18 &  54 35 52 &   180 &   43 &   4.0 \\
        \hline \\[-5pt]
        \end{tabular}
        \end{center}
\end{table}

% created with sort_pht_catalog.pro by Herve Dole on Fri Dec 29 16:32:13 2000

% IFSC /home/hdole/DATA/ISO/firback/catalog/ELAIS_N2/cat2count_elaisn2_aa.tex
 
\begin{table} [!ht]
        \caption{\F\, Catalog in FN2: coordinates are in hours ($\alpha_{2000}$) or degrees ($\delta_{2000}$), minutes, seconds, the flux $S$ and the flux uncertainty $\delta S$ at $170 \mu m $ are in mJy.}
        \label{catalog_elaisn2}
        \begin{center}
        \leavevmode
        \begin{tabular}[h]{cccccc}
                \hline \\[-5pt]
                source       & $\alpha_{2000}$    & $\delta_{2000}$  & $S$  & $\delta S$ & $S/\sigma_c $\\[+5pt]
                \hline \\[-5pt]
FN2\_000 &  16 37 33 &  40 52 26 &  2377 &  213 &  52.8 \\
FN2\_001 &  16 35 08 &  40 59 20 &  1251 &  139 &  27.8 \\
FN2\_002 &  16 36 10 &  41 05 16 &   803 &  102 &  17.8 \\
FN2\_003 &  16 35 25 &  40 55 51 &   682 &   92 &  15.2 \\
FN2\_004 &  16 34 01 &  41 20 49 &   666 &   91 &  14.8 \\
FN2\_005 &  16 32 43 &  41 08 38 &   522 &   78 &  11.6 \\
FN2\_006 &  16 35 06 &  41 10 51 &   346 &   62 &   7.7 \\
FN2\_007 &  16 35 45 &  40 39 14 &   316 &   60 &   7.0 \\
FN2\_008 &  16 35 47 &  41 28 58 &   293 &   58 &   6.5 \\
FN2\_009 &  16 33 55 &  40 53 13 &   291 &   57 &   6.5 \\
FN2\_010 &  16 35 38 &  41 16 58 &   285 &   57 &   6.3 \\
FN2\_011 &  16 38 07 &  40 58 12 &   260 &   55 &   5.8 \\
FN2\_012 &  16 34 13 &  40 56 45 &   249 &   54 &   5.5 \\
FN2\_013 &  16 34 08 &  40 50 52 &   244 &   53 &   5.4 \\
FN2\_014 &  16 38 24 &  41 13 19 &   235 &   52 &   5.2 \\
FN2\_015 &  16 36 07 &  40 55 37 &   223 &   51 &   5.0 \\
FN2\_016 &  16 34 26 &  40 54 07 &   218 &   51 &   4.9 \\
FN2\_017 &  16 34 44 &  41 08 42 &   213 &   50 &   4.7 \\
FN2\_018 &  16 33 38 &  41 01 15 &   212 &   50 &   4.7 \\
FN2\_019 &  16 37 17 &  40 48 36 &   205 &   49 &   4.6 \\
FN2\_020 &  16 32 41 &  41 06 10 &   201 &   49 &   4.5 \\
FN2\_021 &  16 37 58 &  40 51 21 &   196 &   49 &   4.4 \\
FN2\_022 &  16 37 08 &  41 28 26 &   190 &   48 &   4.2 \\
FN2\_023 &  16 33 51 &  40 49 44 &   188 &   48 &   4.2 \\
FN2\_024 &  16 38 56 &  41 02 13 &   185 &   48 &   4.1 \\
FN2\_025 &  16 36 31 &  40 47 38 &   184 &   48 &   4.1 \\
FN2\_026 &  16 36 16 &  40 48 28 &   182 &   47 &   4.0 \\
        \hline \\[-5pt]
        \end{tabular}
        \end{center}
\end{table}

% created with sort_pht_catalog.pro by Herve Dole on Fri Dec 29 16:32:14 2000

%________________________________________________________________
% TABLES COMPLEMENTARY CATALOG
% Input des fichiers tex crees par sort_pht_catalog.pro
%------------------------------------------------------
% CIFSC /home/hdole/DATA/ISO/firback/catalog/Marano/compl_cat2count_marano_aa.tex
 
\begin{table}[!ht]
        \caption{\F\, Complementary Catalog in FSM: coordinates are in hours ($\alpha_{2000}$) or degrees ($\delta_{2000}$), minutes, seconds, the flux $S$ and the flux uncertainty $\delta S$ at $170 \mu m $ are in mJy.}
        \label{compl_catalog_marano}
        \begin{center}
        \leavevmode
        \begin{tabular}[h]{cccccc}
                \hline \\[-5pt]
                source       & $\alpha_{2000}$    & $\delta_{2000}$  & $S$  & $\delta S$ & $S/\sigma_c $\\[+5pt]
                \hline \\[-5pt]
CFSM\_023 &   3 14 06 & -55 16 12 &   173 &   37 &   3.8 \\
CFSM\_024 &   3 07 48 & -55 01 44 &   165 &   37 &   3.7 \\
CFSM\_025 &   3 13 15 & -55 04 44 &   160 &   37 &   3.6 \\
CFSM\_026 &   3 09 43 & -54 43 08 &   160 &   37 &   3.5 \\
CFSM\_027 &   3 13 05 & -55 17 02 &   158 &   37 &   3.5 \\
CFSM\_028 &   3 09 28 & -54 09 57 &   157 &   37 &   3.5 \\
CFSM\_029 &   3 08 38 & -54 57 35 &   157 &   37 &   3.5 \\
CFSM\_030 &   3 10 46 & -54 19 11 &   151 &   36 &   3.4 \\
CFSM\_031 &   3 13 50 & -54 58 15 &   151 &   36 &   3.3 \\
CFSM\_032 &   3 09 41 & -54 21 07 &   149 &   36 &   3.3 \\
CFSM\_033 &   3 12 41 & -54 53 38 &   147 &   36 &   3.3 \\
CFSM\_034 &   3 10 31 & -54 43 51 &   145 &   36 &   3.2 \\
CFSM\_035 &   3 08 09 & -55 09 07 &   142 &   36 &   3.2 \\
CFSM\_036 &   3 11 36 & -54 56 13 &   141 &   36 &   3.1 \\
CFSM\_037 &   3 08 54 & -55 00 46 &   136 &   36 &   3.0 \\
        \hline \\[-5pt]
        \end{tabular}
        \end{center}
\end{table}

% created with sort_pht_catalog.pro by Herve Dole on Fri Dec 29 16:32:13 2000

% CIFSC /home/hdole/DATA/ISO/firback/catalog/ELAIS_N1/compl_cat2count_elaisn1_aa.tex
 
\begin{table}[!ht]
        \caption{\F\, Complementary Catalog in FN1: coordinates are in hours ($\alpha_{2000}$) or degrees ($\delta_{2000}$), minutes, seconds, the flux $S$ and the flux uncertainty $\delta S$ at $170 \mu m $ are in mJy.}
        \label{compl_catalog_elaisn1}
        \begin{center}
        \leavevmode
        \begin{tabular}[h]{cccccc}
                \hline \\[-5pt]
                source       & $\alpha_{2000}$    & $\delta_{2000}$  & $S$  & $\delta S$ & $S/\sigma_c $\\[+5pt]
                \hline \\[-5pt]
CFN1\_056 &  16 11 40 &  54 17 24 &   179 &   43 &   4.0 \\
CFN1\_057 &  16 08 06 &  54 36 36 &   176 &   43 &   3.9 \\
CFN1\_058 &  16 14 18 &  54 06 46 &   175 &   42 &   3.9 \\
CFN1\_059 &  16 08 00 &  54 24 17 &   175 &   42 &   3.9 \\
CFN1\_060 &  16 11 26 &  54 12 54 &   168 &   42 &   3.7 \\
CFN1\_061 &  16 07 49 &  54 41 02 &   168 &   42 &   3.7 \\
CFN1\_062 &  16 14 25 &  54 23 27 &   168 &   42 &   3.7 \\
CFN1\_063 &  16 12 09 &  54 20 13 &   166 &   42 &   3.7 \\
CFN1\_064 &  16 08 21 &  54 38 42 &   166 &   42 &   3.7 \\
CFN1\_065 &  16 14 04 &  54 27 46 &   166 &   42 &   3.7 \\
CFN1\_066 &  16 11 51 &  54 52 37 &   165 &   42 &   3.7 \\
CFN1\_067 &  16 09 44 &  55 01 22 &   165 &   42 &   3.7 \\
CFN1\_068 &  16 10 46 &  54 10 37 &   165 &   42 &   3.7 \\
CFN1\_069 &  16 08 57 &  53 51 54 &   165 &   42 &   3.7 \\
CFN1\_070 &  16 15 42 &  54 32 09 &   164 &   42 &   3.7 \\
CFN1\_071 &  16 09 29 &  54 00 43 &   163 &   42 &   3.6 \\
CFN1\_072 &  16 13 03 &  54 48 32 &   161 &   41 &   3.6 \\
CFN1\_073 &  16 05 37 &  54 31 33 &   161 &   41 &   3.6 \\
CFN1\_074 &  16 08 43 &  53 59 38 &   161 &   41 &   3.6 \\
CFN1\_075 &  16 05 25 &  54 24 57 &   159 &   41 &   3.5 \\
CFN1\_076 &  16 09 53 &  54 03 36 &   159 &   41 &   3.5 \\
CFN1\_077 &  16 07 09 &  54 49 40 &   159 &   41 &   3.5 \\
CFN1\_078 &  16 12 37 &  54 28 51 &   158 &   41 &   3.5 \\
CFN1\_079 &  16 13 29 &  54 43 55 &   157 &   41 &   3.5 \\
CFN1\_080 &  16 10 32 &  55 01 22 &   154 &   41 &   3.4 \\
CFN1\_081 &  16 12 24 &  54 19 22 &   153 &   41 &   3.4 \\
CFN1\_082 &  16 05 39 &  54 41 45 &   151 &   41 &   3.4 \\
CFN1\_083 &  16 10 20 &  54 21 50 &   150 &   41 &   3.3 \\
CFN1\_084 &  16 14 12 &  54 55 58 &   149 &   41 &   3.3 \\
CFN1\_085 &  16 12 58 &  54 45 21 &   149 &   41 &   3.3 \\
CFN1\_086 &  16 10 22 &  54 01 08 &   148 &   40 &   3.3 \\
CFN1\_087 &  16 09 44 &  53 49 37 &   147 &   40 &   3.3 \\
CFN1\_088 &  16 14 34 &  54 38 38 &   146 &   40 &   3.3 \\
CFN1\_089 &  16 12 00 &  54 00 21 &   144 &   40 &   3.2 \\
CFN1\_090 &  16 14 59 &  54 09 17 &   143 &   40 &   3.2 \\
CFN1\_091 &  16 10 44 &  55 06 21 &   141 &   40 &   3.1 \\
CFN1\_092 &  16 10 34 &  54 20 02 &   140 &   40 &   3.1 \\
CFN1\_093 &  16 08 06 &  54 25 37 &   140 &   40 &   3.1 \\
CFN1\_094 &  16 13 52 &  54 28 33 &   138 &   40 &   3.1 \\
CFN1\_095 &  16 08 13 &  55 05 59 &   138 &   40 &   3.1 \\
CFN1\_096 &  16 03 49 &  54 41 49 &   138 &   40 &   3.1 \\
CFN1\_097 &  16 08 38 &  55 04 04 &   137 &   40 &   3.1 \\
CFN1\_098 &  16 11 53 &  54 10 44 &   137 &   40 &   3.0 \\
CFN1\_099 &  16 12 57 &  54 14 38 &   137 &   40 &   3.0 \\
CFN1\_100 &  16 03 42 &  54 49 15 &   136 &   40 &   3.0 \\
CFN1\_101 &  16 09 46 &  54 21 28 &   136 &   40 &   3.0 \\
CFN1\_102 &  16 14 40 &  54 26 34 &   135 &   40 &   3.0 \\
        \hline \\[-5pt]
        \end{tabular}
        \end{center}
\end{table}

% created with sort_pht_catalog.pro by Herve Dole on Fri Dec 29 16:32:13 2000

% CIFSC /home/hdole/DATA/ISO/firback/catalog/ELAIS_N2/compl_cat2count_elaisn2_aa.tex
 
\begin{table}[!ht]
        \caption{\F\, Complementary Catalog in FN2: coordinates are in hours ($\alpha_{2000}$) or degrees ($\delta_{2000}$), minutes, seconds, the flux $S$ and the flux uncertainty $\delta S$ at $170 \mu m $ are in mJy.}
        \label{compl_catalog_elaisn2}
        \begin{center}
        \leavevmode
        \begin{tabular}[h]{cccccc}
                \hline \\[-5pt]
                source       & $\alpha_{2000}$    & $\delta_{2000}$  & $S$  & $\delta S$ & $S/\sigma_c $\\[+5pt]
                \hline \\[-5pt]
CFN2\_027 &  16 37 06 &  41 24 10 &   179 &   47 &   4.0 \\
CFN2\_028 &  16 36 35 &  40 56 06 &   178 &   47 &   4.0 \\
CFN2\_029 &  16 34 20 &  41 06 54 &   178 &   47 &   4.0 \\
CFN2\_030 &  16 35 23 &  40 38 42 &   178 &   47 &   4.0 \\
CFN2\_031 &  16 34 21 &  41 10 19 &   171 &   46 &   3.8 \\
CFN2\_032 &  16 35 03 &  41 31 37 &   168 &   46 &   3.7 \\
CFN2\_033 &  16 34 00 &  41 11 20 &   168 &   46 &   3.7 \\
CFN2\_034 &  16 34 12 &  40 46 26 &   166 &   46 &   3.7 \\
CFN2\_035 &  16 38 50 &  41 05 27 &   166 &   46 &   3.7 \\
CFN2\_036 &  16 37 01 &  40 43 08 &   165 &   46 &   3.7 \\
CFN2\_037 &  16 38 15 &  40 54 25 &   162 &   45 &   3.6 \\
CFN2\_038 &  16 34 32 &  41 22 37 &   161 &   45 &   3.6 \\
CFN2\_039 &  16 36 13 &  40 42 25 &   160 &   45 &   3.6 \\
CFN2\_040 &  16 36 45 &  41 31 22 &   158 &   45 &   3.5 \\
CFN2\_041 &  16 34 31 &  41 00 14 &   156 &   45 &   3.5 \\
CFN2\_042 &  16 35 59 &  40 37 33 &   155 &   45 &   3.4 \\
CFN2\_043 &  16 35 44 &  40 49 26 &   154 &   45 &   3.4 \\
CFN2\_044 &  16 37 26 &  40 45 39 &   150 &   44 &   3.3 \\
CFN2\_045 &  16 34 23 &  41 20 02 &   150 &   44 &   3.3 \\
CFN2\_046 &  16 36 04 &  40 30 21 &   147 &   44 &   3.3 \\
CFN2\_047 &  16 34 51 &  41 20 27 &   147 &   44 &   3.3 \\
CFN2\_048 &  16 37 37 &  40 57 00 &   145 &   44 &   3.2 \\
CFN2\_049 &  16 37 42 &  41 19 11 &   143 &   44 &   3.2 \\
CFN2\_050 &  16 37 18 &  41 16 04 &   142 &   44 &   3.2 \\
CFN2\_051 &  16 36 18 &  41 15 21 &   142 &   44 &   3.2 \\
CFN2\_052 &  16 34 06 &  41 03 10 &   141 &   44 &   3.1 \\
CFN2\_053 &  16 36 23 &  41 23 13 &   138 &   43 &   3.1 \\
CFN2\_054 &  16 36 56 &  41 14 09 &   136 &   43 &   3.0 \\
        \hline \\[-5pt]
        \end{tabular}
        \end{center}
\end{table}

% created with sort_pht_catalog.pro by Herve Dole on Fri Dec 29 16:32:14 2000

%______________________________________________________________

\end{document}